\documentclass[12pt, jmp]{revtex4}

\usepackage{graphicx}
\usepackage{dcolumn}
\usepackage{bm}

\usepackage{amsmath}
\usepackage{amsxtra}
\usepackage{amscd}
\usepackage{amsthm}
\usepackage{amsfonts}
\usepackage{amssymb}
\usepackage{eucal}

\input prepictex
\input pictex
\input postpictex
\usepackage{mathptm}

\newtheorem{theorem}{Theorem}[section]
\newtheorem{lemma}[theorem]{Lemma}

\newtheorem{proposition}[theorem]{Proposition}
\theoremstyle{definition}
\newtheorem{definition}[theorem]{Definition}

\theoremstyle{remark}
\newtheorem{remark}[theorem]{Remark}

\numberwithin{equation}{section}

\newcommand{\R}{{\mathbb R}}
\newcommand{\C}{{\mathbb C}}

\newcommand{\barh}{{\bar h}}
\newcommand{\cA}{{\mathcal A}}
\newcommand{\cG}{{\mathcal G}}
\newcommand{\gphi}{{}^{\phi}{g}}

\begin{document}


\title{Riemannian Geometry of Noncommutative Surfaces}

\author{M. Chaichian}
\email{Masud.Chaichian@helsinki.fi}
\author{A. Tureanu}
\email{Anca.Tureanu@helsinki.fi}
\affiliation{Department of Physics, University of Helsinki\\
and Helsinki Institute of Physics, P.O. Box 64, 00014 Helsinki,
Finland}

\author{R. B. Zhang}
\email{rzhang@maths.usyd.edu.au}
\affiliation{School of Mathematics
and Statistics, University of Sydney, Sydney, NSW 2006, Australia}

\author{Xiao  Zhang}
\email{xzhang@amss.ac.cn}
\affiliation{Institute of Mathematics,
Academy of Mathematics and Systems Science, Chinese Academy of
Sciences, Beijing 100190, China}

\begin{abstract}
A Riemannian geometry of noncommutative $n$-dimensional surfaces is
developed as a first step towards the construction of a consistent
noncommutative gravitational theory. Historically, as well,
Riemannian geometry was recognized to be the underlying structure of
Einstein's theory of general relativity and led to further
developments of the latter. The notions of metric and connections on
such noncommutative surfaces are introduced and it is shown that the
connections are metric-compatible, giving rise to the corresponding
Riemann curvature. The latter also satisfies the noncommutative
analogue of the first and second Bianchi identities. As examples,
noncommutative analogues of the sphere, torus and hyperboloid are
studied in detail. The problem of covariance under appropriately
defined general coordinate transformations is also discussed and
commented on as compared with other treatments.
\end{abstract}
\maketitle

%
%
\section{Introduction}\label{introd}

In recent years there has been much progress in developing theories
of noncommutative geometry and exploring their applications in
physics. Many viewpoints were adopted and different mathematical
approaches were followed by different researchers. Connes' theory
\cite{Co} (see also Ref. \cite{GVF}) formulated within the
framework of $C^*$-algebras is the most successful, which
incorporates cyclic cohomology and K-theory, and gives rise to
noncommutative versions of index theorems. Theories generalizing
aspects of algebraic geometry were also developed (see, e.g., Ref.
\cite{SV} for a review and references). A notion of
noncommutative schemes was formulated, which seems to provide a
useful framework for developing noncommutative algebraic geometry.

A major advance in theoretical physics in recent years was the
deformation quantization of Poisson manifolds by Kontsevich (see
Ref. \cite{Ko} for the final form of this work). This sparked
intensive activities investigating applications of noncommutative
geometries to quantum theory. Seiberg and Witten \cite{SW} showed
that the anti-symmetric tensor field arising from massless states of
strings can be described by the noncommutativity of a spacetime,
\begin{equation}\label{cr}
[x^\mu,x^\nu]_\ast=i\theta^{\mu\nu},\ \ \theta^{\mu\nu}\
\mbox{constant matrix},
\end{equation}
where the multiplication of the algebra of functions is governed by
the Moyal product
\begin{equation}\label{moyal}
(f\ast
g)(x)=f(x)\exp{\left(\frac{i}{2}\theta^{\mu\nu}
\overleftarrow{\partial}_\mu\overrightarrow{\partial}_\nu\right)}g(x)\,.
\end{equation}
A considerable amount of research was done both prior and after Ref.
\cite{SW}, and we refer to Refs. \cite{V, DN} for
reviews and references.

An earlier and independent work is a seminal paper \cite{DFR} by
Doplicher, Fredenhagen and Roberts, which laid down the fundamentals
of quantum field theory on noncommutative spacetime. These authors
started with a theoretical examination of the long held belief by
the physics community that the usual notion of spacetime needed to
be modified at the Planck scale, and convincingly demonstrated that
spacetime becomes noncommutative in that the coordinates describing
spacetime points become operators similar to those in quantum
mechanics. Therefore, noncommutative geometry is indeed a way to
describe physics at Planck scale.

A consistent formulation of a noncommutative version of general
relativity could give insight into a gravitational theory compatible
with quantum mechanics. A unification of general relativity with
quantum mechanics has long been sought after but remains as elusive
as ever despite the extraordinary progress in string theory for the
last two decades. The noncommutative geometrical approach to gravity
could provide an alternative route. Much work has already been done
in this general direction, see, e.g., Refs. \cite{C01, C04,
MM, M05, ADMW1, ADMW2, BGMZ} and references therein. In particular,
different forms of noncommutative Riemannian geometries were
proposed \cite{MM, M05, ADMW1, ADMW2}, which retain some of the
familiar geometric notions like metric and curvature. Noncommutative
analogues of the Hilbert-Einstein action were also suggested
\cite{C01, C04, ADMW1, ADMW2} by treating noncommutative gravity as
gauge theories.

The noncommutative spacetime with the Heisenberg-like commutation
relation (\ref{cr}) violates Lorentz symmetry but
%
%
was shown \cite{CKNT, CKNT1} to have a quantum symmetry under the
twisted Poincar\'e algebra. The Abelian twist element
\begin{equation}\label{twist}
{\mathcal
F}=\exp\left(-\frac{i}{2}\theta^{\mu\nu}\partial_\mu\otimes\partial_\nu\right)
\end{equation}
was used in Ref. \cite{CKNT, CKNT1} to twist the universal
enveloping algebra of the Poincar\'e algebra, obtaining a
noncommutative multiplication for the algebra of functions on the
Poincar\'e group closely related to the Moyal product (\ref{moyal}).
It is then natural to try to extend the procedure to other
symmetries of noncommutative field theory and investigate whether
the concept of twist provides a new symmetry principle for
noncommutative spacetime.

The same Abelian twist element (\ref{twist}) was used in Refs.
\cite{ADMW1, ADMW2} for deforming the algebra of
diffeomorphisms when attempting to obtain general coordinate
transformations on the noncommutative space-time. It is interesting
that Refs. \cite{ADMW1, ADMW2} proposed gravitational theories
which are different from the low energy limit of strings \cite{AMV}.
However, based on physical arguments one would expect the Moyal
product to be frame-dependent and transform under the general
coordinate transformation. If the twist element is chosen as
(\ref{twist}), the Moyal $\ast$-product is fixed once for all. This
is likely to lead to problems similar to those observed in Ref.
\cite{CT} when one attempted to deform the internal gauge
transformations with the same twist element \eqref{twist}.
Nevertheless twisting is expected to be a productive approach to the
formulation of a noncommutative gravitational theory when
implemented consistently. A ``covariant twist" was proposed for
internal gauge transformations in Ref. \cite{CTZ}, but it
turned out that the corresponding star-product would not be
associative.


Works on the noncommutative geometrical approach to gravity may be
broadly divided into two types. One type attempts to develop
noncommutative versions of Riemannian geometry axiomatically (that
is, formally), while the other adapts general relativity to the
noncommutative setting in an intuitive way. A problem is the lack of
any safeguard against mathematical inconsistency in the latter type
of works, and the same problem persists in the first type of works
as well, since it is not clear whether nontrivial examples exist
which satisfy all the axioms of the formally defined theories.

The aim of the present paper is to develop a theory of
noncommutative Riemannian geometry by extracting an axiomatic
framework from highly nontrivial and transparently consistent
examples. Our approach is mathematically different from that of
Refs. \cite{ADMW1, ADMW2} and also quite far removed from the
quantum group theoretical noncommutative Riemannian geometry
\cite{M05} (see also references in Ref. \cite{M05} and
subsequent publications by the same author).

Recall that 2-dimensional surfaces embedded in the Euclidean 3-space
provide the simplest yet nontrivial examples of Riemannian geometry.
The Euclidean metric of the 3-space induces a natural metric for a
surface through the embedding; the Levi-Civita connection and the
curvature of the tangent bundle of the surface can thus be described
explicitly (for the theory of surfaces, see, e.g., the textbook Ref.
\cite{doC}).  As a matter of fact, Riemannian geometry
originated from Gauss' work on surfaces embedded in $3$-dimensional
Euclidean space.

More generally, Whitney's theorem enables the embedding of any
smooth Riemannian manifold as a high dimensional surface in a flat
Euclidean space of high enough dimension (see, e.g., Theorem 9 and
Theorem 11.1.1 in Ref. \cite{DFN}). The embedding also allows
transparent construction and interpretation of all structures
related to the Riemannian metric of $M$ as in the 2-dimensional
case.

This paper develops noncommutative deformations of Riemannian
geometry in the light of Whitney's theorem. The first step is to
deform the algebra of functions on a domain of the Euclidean space.
We begin Section \ref{surfaces} by introducing the Moyal algebra
$\cA$, which is a noncommutative deformation \cite{Ge} of the
algebra of smooth functions on a region of $\R^2$. The rest of
Section \ref{surfaces} develops a noncommutative Riemannian geometry
for noncommutative analogues of 2-dimensional surfaces embedded in
3-space. Working over the Moyal algebra $\cA$, we show that much of
the classical differential geometry for surfaces generalizes {\em
naturally} to this noncommutative setting. In Section
\ref{examples}, three illuminating examples are constructed, which
are respectively noncommutative analogues of the sphere, torus and
hyperboloid. Their noncommutative geometries are studied in detail.

We emphasize that the embeddings play a crucial role in our current
understanding of the geometry of the 2-dimensional noncommutative
surfaces.  The metric of a noncommutative surface is constructed in
terms of the embedding; the necessity of a left connection and also
a right connection then naturally arises; even the definition of the
curvature tensor is forced upon us by the context. Indeed, the extra
information obtained by considering embeddings provides the guiding
principles, which are lacking up to now, for building a theory of
noncommutative Riemannian geometry.

Once the noncommutative Riemannian geometry of the 2-dimensional
surfaces is sorted out, its generalization to the noncommutative
geometries corresponding to $n$-dimensional surfaces embedded in
spaces of higher dimensions is straightforward. This is discussed in
Section \ref{general}.

Recall that the basic principle of general relativity is general
covariance. We study in Section \ref{transformations} general
coordinate transformations for noncommutative surfaces, which are
brought about by gauge transformations on the underlying
noncommutative associative algebra $\cA$ (over which noncommutative
geometry is constructed).  A new feature here is that the general
coordinate transformations affect the multiplication of the
underlying associative algebra $\cA$ as well, turning it into
another algebra nontrivially isomorphic to $\cA$. We make comparison
with classical Riemannian geometry, showing that the gauge
transformations should be considered as noncommutative analogues of
diffeomorphisms.

The theory of surfaces developed over the deformation of the algebra
of smooth functions on some region in $\R^n$ now suggests a general
theory of noncommutative Riemannian geometry of $n$-dimensional
surfaces over arbitrary unital associative algebras with
derivations. We present an outline of this general theory in Section
\ref{generalization}.

We conclude this section with a remark on the presentation of the
paper. As indicated above, we start from the simplest nontrivial
examples of noncommutative Riemannian geometries and gradually
extend the results to build up a theory of generality. This
``experimental approach" is not the optimal format for presenting
mathematics, as all special cases repeat the same pattern. However,
it has the distinctive advantage that the general theory obtained in
this way stands on firm ground.

\section{Noncommutative surfaces and their embeddings}\label{surfaces}
\subsection{Noncommutative surfaces and their embeddings}

The first step in constructing noncommutative deformations of
Riemannian geometry is the deformation of algebras of functions. Let
us fix a region $U$ in $\R^2$, and write the coordinate of a point
$t$ in $U$ as $(t_1, t_2)$.  Let  $\barh$ be a real indeterminate,
and denote by $\R[[\barh]]$ the ring of formal power series in
$\barh$. Let $\cA$ be the set of the formal power series in $\barh$
with coefficients being real smooth functions on $U$. Namely, every
element of $\cA$ is of the form $\sum_{i\ge 0} f_i\barh^i$ where
$f_i$ are smooth functions on $U$. Then $\cA$ is an
$\R[[\barh]]$-module in the obvious way.

Given any two smooth functions $f$ and $g$ on $U$, we denote by $f
g$ the usual point-wise product of the two functions. We also define
their star-product (or more precisely, Moyal product) by
\begin{eqnarray}\label{product}
f\ast g = \lim_{t'\rightarrow t}\,  \exp{\left[\barh \left(
\frac{\partial}{\partial t_1} \frac{\partial}{\partial t'_2} -
\frac{\partial}{\partial t_2} \frac{\partial}{\partial
t'_1}\right)\right]}f(t) g(t'),
\end{eqnarray}
where the exponential $\exp[\barh ( \frac{\partial}{\partial t_1}
\frac{\partial}{\partial t'_2} - \frac{\partial}{\partial t_2}
\frac{\partial}{\partial t'_1})]$ is to be understood as a power
series in the differential operator $\frac{\partial}{\partial t_1}
\frac{\partial}{\partial t'_2} - \frac{\partial}{\partial t_2}
\frac{\partial}{\partial t'_1}$. More explicitly, let
\begin{eqnarray}\label{mup}
\mu_p: \cA/\barh\cA\otimes\cA/\barh\cA\longrightarrow\cA/\barh\cA,
\quad p=0, 1, 2, \dots,
\end{eqnarray}
be $\R$-linear maps defined by
\[
\mu_p(f, g) = \lim_{t'\rightarrow t} \frac{1}{p!}\left(
\frac{\partial}{\partial t_1} \frac{\partial}{\partial t'_2} -
\frac{\partial}{\partial t_2} \frac{\partial}{\partial
t'_1}\right)^p f(t) g(t').
\]
Then $f\ast g = \sum_{p=0}^\infty \barh^p \mu_p(f, g)$. It is
evident that $f \ast g$ lies in $\cA$. We extend this star-product
$\R[[\barh]]$-linearly to all elements in $\cA$ by letting
\begin{eqnarray*}
(\sum f_i \barh^i)\ast (\sum g_j \barh^j) :=\sum f_i\ast g_j
\barh^{i+j}.
\end{eqnarray*}
It has been known since the early days of quantum mechanics that the
Moyal product is associative (see, e.g., Ref. \cite{Ko} for a
reference), thus we arrive at an associative algebra over
$\R[[\barh]]$, which is a deformation \cite{Ge} of the algebra of
smooth functions on $U$. We shall usually denote this associative
algebra by $\cA$, but when it is necessary to make explicit the
multiplication of the algebra, we shall write it as $(\cA, \ast)$.

\begin{remark}
For the sake of being explicit, we restrict ourselves to consider
the Moyal product (defined by \eqref{product}) only in this section.
As we shall see in Sections \ref{generalization} and \ref{transformations},
the theory of noncommutative surfaces to be
developed in this paper extends to more general star-products over
algebras of smooth functions.
\end{remark}

Write $\partial_i$ for $\frac{\partial}{\partial t_i}$, and extend
$\R[[\barh]]$-linearly the operators $\partial_i$ to $\cA$. One can
easily verify that for smooth functions $f$ and $g$,
\begin{eqnarray}\label{Leibniz}
\partial_i(f\ast g)  =(\partial_i f)\ast g + f\ast (\partial_i g),
\end{eqnarray}
that is, the operators $\partial_i$ are derivations of the algebra
$\cA$.

Let $\cA^3=\cA\oplus\cA\oplus \cA$. There is a natural two-sided
$\cA$-module structure on $\cA^3\otimes_{\R[[\barh]]}\cA^3$,
defined for all $a, b\in\cA$, and $X\otimes Y\in
\cA^3\otimes_{\R[[\barh]]}\cA^3$ by $a\ast(X\otimes Y)\ast b =
a\ast X\otimes Y\ast b$. Define the map
\begin{eqnarray}\label{dotproduct}
\cA^3\otimes_{\R[[\barh]]}\cA^3 \longrightarrow \cA, & & (a, b,
c)\otimes (f, g, h) \mapsto a\ast f + b\ast g + c\ast h,
\end{eqnarray}
and denote it by $\bullet$. This is a map of two-sided $\cA$-modules
in the sense that for any $X, Y\in \cA^3$ and $a, b\in \cA$, $(a\ast
X)\bullet (Y\ast b) = a\ast (X\bullet Y)\ast b$. We shall refer to
this map as the {\em dot-product}.

Let $X=(X^1, X^2, X^3)$ be an element of $\cA^3$, where the
superscripts of $X^1$, $X^2$ and $X^3$ are not powers but are
indices used to label the components of a vector as in the usual
convention in differential geometry. We set $\partial_i X=
(\partial_i X^1,
\partial_i X^2, \partial_i X^3)$, and define the following $2\times
2$-matrix over $\cA$
\begin{eqnarray}\label{metric}
g=\begin{pmatrix}g_{1 1} & g_{1 2}\\
g_{2 1} & g_{2 2}\end{pmatrix},  &\quad& g_{i j} = \partial_i
X\bullet
\partial_j X.
\end{eqnarray}
Let $g^0 = g\mod\barh$, which is a $2\times 2$-matrix of smooth
functions on $U$.
\begin{definition}
We call an element $X\in \cA^3$ (the {\em noncommutative embedding}
in $\cA^3$ of) a {\em noncommutative surface} if $g^0$ is invertible
for all $t\in U$. In this case, we call $g$ the {\em metric} of the
noncommutative surface.
\end{definition}

Given a noncommutative surface $X$ with a metric $g$, there exists a
unique $2\times 2$-matrix $\left(g^{i j}\right)$ over $\cA$ which is
the right inverse of $g$, i.e.,
\[
g_{i j}\ast g^{j k} =\delta_i^k,
\]
where we have used Einstein's convention of summing
over repeated indices. To see the existence of the right inverse, we
write $g_{i j} =\sum_p \barh^p g_{i j}[p]$ and $g^{i j} =\sum_p
\barh^p \tilde g^{i j}[p]$, where $(g^{i j}[0])$ is the inverse of
$(g_{i j}[0])$. Now in terms of the maps $\mu_k$ defined by
\eqref{mup}, we have
\begin{eqnarray*}
\delta_i^k &=g_{i j}\ast g^{j k} &= \sum_q \barh^q
\sum_{m+n+p=q}\mu_p(g_{i j}[m], g^{j k}[n]),
\end{eqnarray*}
which is equivalent to
\begin{eqnarray*}
g^{i j}[q] &=& -\sum_{n=1}^q\sum_{m=0}^{q-n}g^{i k}[0] \mu_n(g_{k
l}[m], \, g^{l j}[q-n-m]).
\end{eqnarray*}
Since the right-hand side involves only $g^{l j}[r]$ with $r<q$,
this equation gives a recursive formula for the right inverse of
$g$.

In the same way, we can also show that there also exists a unique
left inverse of $g$. It follows from the associativity of
multiplication of matrices over any associative algebra that the
left and right inverses of $g$ are equal.

\begin{definition}
Given a noncommutative surface $X$, let
\[
E_i=\partial_i X, \quad i=1, 2,
\]
and call the left $\cA$-module
$TX$ and right $\cA$-module $\tilde TX$ defined by
\[
T X=\{a\ast E_1+ b\ast E_2\mid a, b\in \cA \}, \quad  \tilde T
X=\{E_1 \ast a +E_2\ast b\mid a, b\in \cA \}
\]
the {\em left} and {\em right tangent bundles} of the
noncommutative surface respectively.
\end{definition}
Then $TX\otimes_{\R[[\barh]]} \tilde TX $ is a two-sided
$\cA$-module.
\begin{proposition}
The metric induces a homomorphism of two-sided $\cA$-modules
\[
g: TX\otimes_{\R[[\barh]]} \tilde TX \longrightarrow \cA,
\]
defined for any $Z=z^i\ast E_i \in TX$ and $W= E_i\ast w^i\in \tilde
TX$ by
\[
Z\otimes W \mapsto g(Z, W)=z^i\ast g_{i j}\ast w^j.
\]
\end{proposition}
It is easy to see that the map is indeed a homomorphism of two-sided
$\cA$-modules, and it clearly coincides with the restriction of the
dot-product to $T X\otimes_{\R[[\barh]]} \tilde T X$.

Since the metric $g$ is invertible, we can define
\begin{eqnarray}\label{E-upper}
E^i = g^{i j}\ast E_j, \quad \tilde E^i = E_j \ast g^{j i},
\end{eqnarray}
which belong to $TX$ and $\tilde TX$ respectively. Then
\[
g(E^i, E_j)= \delta^i_j=g(E_j, \tilde E^i), \quad g(E^i, \tilde E^j)= g^{i
j}.
\]

Now any $Y\in \cA^3$ can be written as $Y= y^i \ast E_i + Y^\bot$,
with $y^i = Y\bullet \tilde E^i$, and $Y^\bot=Y-y^i \ast E_i$. We
shall call $y^i \ast E_i$ the {\em left tangential component} and
$Y^\bot$ the {\em left normal component} of $Y$. Let
\[
(T X)^\bot=\{N\in \cA^3 \mid
N\bullet E_i=0, \forall i\},
\]
which is clearly a left $\cA$-submodule of $\cA^3$. In a similar
way, we may also decompose $Y$ into $Y = E_i\ast \tilde y^i +
\tilde Y^\bot$ with the {\em right tangential component} of $Y$
given by $\tilde y^i = E^i\bullet Y$ and  the {\em right normal
component} by $\tilde Y^\bot=Y-E_i\ast \tilde y^i$. Let
\[
(\tilde T X)^\bot=\{N\in \cA^3 \mid E_i\bullet N=0, \forall i\},
\]
which is a right $\cA$-submodule of $\cA^3$. Therefore, we have the
following decompositions
\begin{eqnarray}\label{decomposition}
\begin{aligned}
\cA^3 &=& T X \oplus (T X)^\bot,  &\quad  \text{as left $\cA$-module}, \\
\cA^3 &=& \tilde T X \oplus (\tilde T X)^\bot, &\quad \text{as right
$\cA$-module}.
\end{aligned}
\end{eqnarray}

It follows that the tangent bundles are finitely generated
projective modules over $\cA$. Following the general philosophy of
noncommutative geometry \cite{Co}, we may regard finitely
generated projective modules over $\cA$ as vector bundles on the
noncommutative surface. This justifies the terminology of left and
right tangent bundles for $TX$ and $\tilde TX$.

In fact $TX$ and $\tilde TX$ are free left and right $\cA$-modules
respectively, as $E_1$ and $E_2$ form $\cA$-bases for them. Consider
$TX$ for example. If there exists a relation $a^i\ast E_i =0$, where
$a^i\in\cA$, we have $a^i\ast E_i\bullet E_j$ $= a^i\ast g_{i j}$
$=0$, $\forall j$. The invertibility of the metric then leads to
$a^i=0$, $\forall i$. Since $E_1$ and $E_2$ generate $TX$, they
indeed form an $\cA$-basis of $TX$.

One can introduce connections to the tangent bundles by following
the standard procedure in the theory of surfaces \cite{doC}.

\begin{definition}\label{nabla}
Define operators
\[\nabla_i: T X\longrightarrow T X, \quad i=1, 2,\]
by requiring that $\nabla_i Z$ be equal to the left tangential
component of $\partial_i Z$ for all $Z\in TX$. Similarly define
\[\tilde\nabla_i: \tilde T X\longrightarrow \tilde T X, \quad i=1, 2,\]
by requiring that $\tilde\nabla_i \tilde Z$ be equal to the right
tangential component of $\partial_i \tilde Z$ for all $\tilde Z\in
\tilde T X$. Call the set consisting of the operators $\nabla_i$
(respectively $\tilde\nabla_i$) a {\em connection} on $TX$
(respectively $\tilde TX$).
\end{definition}

The following result justifies the terminology.
\begin{lemma}
For all $Z\in TX$,  $W\in \tilde TX$ and $f\in \cA$,
\begin{eqnarray}\label{linear}
\nabla_i(f\ast Z)= \partial_i f \ast Z+ f\ast\nabla_i Z, \quad
\tilde\nabla_i(W\ast f)= W \ast  \partial_i f+ \tilde \nabla_i W
\ast f.
\end{eqnarray}
\end{lemma}
\begin{proof}
By the Leibniz rule \eqref{Leibniz} for $\partial_i$,
\[
\partial_i(f\ast Z)=
(\partial_i f)\ast Z+ f\ast(\partial_i Z), \quad
\partial_i(W\ast
f)= W\ast(\partial_i f)+ W\ast(\partial_i f).
\]
The lemma immediately follows from the tangential components of
these relations under the decompositions \eqref{decomposition}.
\end{proof}

In order to describe the connections more explicitly, we note that
there exist $\Gamma_{i j}^k$ and $\tilde\Gamma_{i j}^k$ in $\cA$
such that
\begin{eqnarray}\label{Gamma}
\nabla_i E_j= \Gamma_{i j}^k \ast E_k, && \tilde \nabla_i E_j= E_k
\ast \tilde \Gamma_{i j}^k.
\end{eqnarray}
Because the metric is invertible, the elements $\Gamma_{i j}^k$ and
$\tilde \Gamma_{i j}^k$ are uniquely defined by equation
\eqref{Gamma}. We have
\begin{eqnarray}\label{Gamma1}
\Gamma_{i j}^k =  \partial_i E_j\bullet \tilde E^k &\quad&
\tilde\Gamma_{i j}^k = E^k \bullet  \partial_i E_j.
\end{eqnarray}
It is evident that $\Gamma_{i j}^k$ and $\tilde \Gamma_{i j}^k$ are
symmetric in the indices $i$ and $j$.  The following closely related
objects will also be useful later:
\[
\Gamma_{i j k} =  \partial_i E_j\bullet E_k , \quad
\tilde\Gamma_{i j k} = E_k \bullet  \partial_i E_j.
\]
In contrast to the commutative case, $\Gamma_{i j}^k$ and
$\tilde\Gamma_{i j}^k$ do not coincide in general. We have
\[
\Gamma_{i j}^k =  {}_c\Gamma_{i j l}\ast g^{l k} + \Upsilon_{i j
 l}\ast g^{l k}, \quad \tilde\Gamma_{i j}^k = g^{k l}\ast
{}_c\Gamma_{i j l} - g^{k l}\ast \Upsilon_{i j l},
\]
where
\begin{eqnarray*}
\begin{aligned}
_c\Gamma_{i j l} &=& \frac{1}{2} \left(\partial_i g_{j l} +
\partial_j g_{l i}
-\partial_l g_{j i} \right), \\
\Upsilon_{i j l} &=& \frac{1}{2} \left(\partial_iE_j\bullet E_l -
E_l\bullet \partial_i E_j\right).
\end{aligned}
\end{eqnarray*}
We shall call $\Upsilon_{i j l}$ the {\em noncommutative torsion} of
the noncommutative surface. Therefore the left and right connections
involve two parts. The part $ _c\Gamma_{i j l}$ depends on the
metric only, while the noncommutative torsion embodies extra
information. For a noncommutative surface embedded in $\cA^3$, the
noncommutative torsion depends explicitly on the embedding. In the
classical limit with $\barh=0$, $\Upsilon_{i j}^k$ vanishes and both
$\Gamma_{i j}^k$ and $\tilde\Gamma_{i j}^k$ reduce to the standard
Levi-Civita connection.

We have the following result.
\begin{proposition}\label{prop:compatible}
The connections are metric compatible in the following sense
\begin{eqnarray}\label{compatible}
\partial_i g(Z, \tilde Z)=g(\nabla_i Z, \tilde Z) + g(Z, \tilde\nabla_i\tilde
Z), &\quad& \forall Z\in TX, \ \tilde Z\in \tilde TX.
\end{eqnarray}
This is equivalent to the fact that
\begin{eqnarray}\label{compatible1}
\partial_i g_{j k} -\Gamma_{i j k} - \tilde\Gamma_{i k j} =0.
\end{eqnarray}
\end{proposition}
\begin{proof} Since $g$ is a map of two-sided $\cA$-modules, it
suffices to prove \eqref{compatible} by verifying the special case
with $Z=E_j$ and $\tilde Z=E_k$. We have
\begin{eqnarray*}\partial_i g(E_j, E_k)&=&\partial_i (E_j\bullet E_k)
\\
&=& \partial_i E_j\bullet E_k + E_j\bullet  \partial_i E_k\\
&=& g(\nabla_i E_j,  E_k) + g(E_j,  \tilde \nabla_i E_k),
\end{eqnarray*}
where the second equality is equivalent \eqref{compatible1}. This
proves both statements of the proposition.
\end{proof}

\begin{remark}\label{Gammas}
In contrast to the commutative case, equation \eqref{compatible1} by
itself is not sufficient to uniquely determine the connections
$\Gamma_{i j k}$ and $\tilde\Gamma_{i j k}$; the noncommutative
torsion needs to be specified independently. This is similar to the
situation in supergeometry, where torsion is determined by other
considerations.
\end{remark}

At this point we should relate to the literature. The metric
introduced here resembles similar notions in Refs. \cite{MM,
DM, DMMM, DHLS}; also our left and right connections and their
metric compatibility have much similarity with Definitions 2 and 3
in Ref. \cite{DHLS}. However, there are crucial differences.
Our left (respectively right) tangent bundle is a left (respectively
right) $\cA$-module only, while in Refs. \cite{DM, DHLS} there
is only one ``tangent bundle" $T$ which is a bimodule over some
algebra (or Hopf algebra) $B$. The metrics defined in Refs.
\cite{MM, DM, DMMM, DHLS} are maps from $T\otimes_B T$ to $B$.

\begin{remark}
A noteworthy feature of the metric in Ref. \cite{MM} is that a
particular moving frame can be chosen to make all the components of
the metric central (see equation (3.22) in Ref. \cite{MM}). In
the context of the Moyal algebra, this amounts to that the metric is
a constant matrix.
\end{remark}

\subsection{Curvatures and second fundamental form}

Let $[\nabla_i, \nabla_j]:=\nabla_i \nabla_j- \nabla_j \nabla_i$ and
${[\tilde\nabla_i,  \tilde\nabla_j]}:= \tilde\nabla_i
\tilde\nabla_j- \tilde\nabla_j \tilde\nabla_i$. Straightforward
calculations show that for all $f\in \cA$,
\begin{eqnarray*}
\begin{aligned}
&{[}\nabla_i, \nabla_j{]}
(f\ast Z) =f\ast[\nabla_i, \nabla_j]Z,   &Z\in TX,  \\
&{[}\tilde\nabla_i,  \tilde\nabla_j{]}(W\ast f) = [\tilde\nabla_i,
\tilde\nabla_j]W\ast f, & W\in \tilde TX.
\end{aligned}
\end{eqnarray*}
Clearly the right-hand side of the first equation belongs to $TX$,
while that of the second equation belongs to $\tilde TX$.  We
re-state these important facts as a proposition.
\begin{proposition}\label{prop:curvature}
The following maps
\begin{eqnarray*}
[\nabla_i, \nabla_j]: TX \longrightarrow TX, && {[\tilde\nabla_i,
\tilde\nabla_j]}: \tilde TX \longrightarrow \tilde TX
\end{eqnarray*}
are left and right $\cA$-module homomorphisms respectively.
\end{proposition}

Since $TX$ (respectively $\tilde TX$) is generated by $E_1$ and
$E_2$ as a left (respectively right) $\cA$-module, by Proposition
\ref{prop:curvature}, we can always write
\begin{eqnarray}\label{curvature}
{[\nabla_i, \nabla_j]} E_k = R_{k i j}^l\ast E_l, & \quad &
{[}\tilde\nabla_i, \tilde\nabla_j{]}E_k = E_l\ast \tilde R_{k i
j}^l
\end{eqnarray}
for some $R_{k i j}^l, \tilde R_{k i j}^l\in\cA$.
\begin{definition}
We refer to $R_{k i j}^l$ and $\tilde R_{k i j}^l$ respectively as
the {\em Riemann curvatures} of the left and right tangent bundles
of the noncommutative surface $X$.
\end{definition}

The Riemann curvatures are uniquely determine by the relations
\eqref{curvature}. In fact, we have
\begin{eqnarray}
\begin{aligned}
R_{k i j}^l  &=&  g({[\nabla_i, \nabla_j]} E_k, \tilde E^l),
&\quad& \tilde R_{k i j}^l  &=&  g(E^l, {[}\tilde\nabla_i,
\tilde\nabla_j{]}E_k).
\end{aligned}
\end{eqnarray}
Simple calculations yield the following result.
\begin{lemma}\label{curvature1}
\[
\begin{aligned} R_{k i j}^l &=& -\partial_j\Gamma_{i k}^l -
\Gamma_{i k}^p \ast \Gamma_{j p}^l + \partial_i\Gamma_{j k}^l
+\Gamma_{j k}^p\ast \Gamma_{i p}^l ,\\
\tilde R_{k i j}^l &=& -\partial_j\tilde\Gamma_{i k}^l -
\tilde\Gamma_{j p}^l\ast\tilde\Gamma_{i k}^p
+\partial_i\tilde\Gamma_{j k}^l + \tilde\Gamma_{i p}^l\ast
\tilde\Gamma_{j k}^p
\end{aligned}
\]
\end{lemma}
\begin{proposition}\label{curvature3}
Let $R_{l k i j} = R_{k i j}^p\ast g_{p l}$ and $\tilde R_{l k i j}
= - g_{k p}\ast \tilde R_{l i j}^p$. The Riemann curvatures of the
left and right tangent bundles coincide in the sense that $R_{k l i
j}=\tilde R_{ k l i j}$.
\end{proposition}
\begin{proof}
By Proposition \ref{prop:compatible}, we have $R_{l k i j }=
(\nabla_i \nabla_j- \nabla_j \nabla_i) E_k\bullet E_l$, which can
be re-written as
\begin{eqnarray*}
R_{l k i j } &=&\partial_i (\nabla_j E_k\bullet E_l)-\nabla_j
E_k\bullet \tilde\nabla_iE_l\\
&&  -\partial_j (\nabla_i E_k\bullet E_l)+ \nabla_i
E_k\bullet\tilde\nabla_j E_l \\
&=& \partial_i (\nabla_j E_k\bullet E_l + E_k\bullet\tilde\nabla_j
E_l) -
E_k\bullet\tilde\nabla_i\tilde\nabla_j E_l\\
&&-\partial_j (\nabla_i E_k\bullet E_l + E_k\bullet \tilde\nabla_i
E_l) + E_k\bullet\tilde\nabla_j \tilde\nabla_i E_l.
\end{eqnarray*}
Again by Proposition \ref{prop:compatible}, the first term on the
far right-hand side can be written as $\partial_i\partial_j g_{k
l}$, and the third term can be written as $-\partial_i\partial_j
g_{k l}$. Thus they cancel out, and we arrive at
\begin{eqnarray*}
R_{l k i j }=&-
E_k\bullet(\tilde\nabla_i\tilde\nabla_j-\tilde\nabla_j\tilde\nabla_i)E_l
=&\tilde R_{l k i j}.
\end{eqnarray*}
\end{proof}
Because of the proposition, we only need to study the Riemannian
curvature on one of the tangent bundles. Note that $R_{ k l i
j}=-R_{k l j i }$, but there is no simple rule to relate $R_{l k i j
}$ to $R_{k l i j}$ in contrast to the commutative case.

\begin{definition}
Let
\begin{eqnarray}
R_{i j} = R^p_{i p j}, &\quad& R= g^{j i}\ast R_{i j},
\end{eqnarray}
and call them the {\em Ricci curvature} and {\em scalar curvature}
of the noncommutative surface respectively.
\end{definition}
Then obviously
\begin{eqnarray}\label{Ricci}
R_{i j} = -g([\nabla_j, \nabla_l]E_i, \tilde E^l), \quad R=
-g([\nabla_i, \nabla_k] E^i, \tilde E^k).
\end{eqnarray}

In the theory of classical surfaces, the second fundamental form
plays an important role. A similar notion exists for noncommutative
surfaces.
\begin{definition}\label{sff}
We define the left and right {\em second fundamental forms} of the
noncommutative surface $X$ by
\begin{eqnarray}\label{h}
h_{i j} = \partial_i E_j -\Gamma_{i j}^k \ast E_k, && \tilde h_{i
j}= \partial_i E_j - E_k \ast \tilde \Gamma_{i j}^k.
\end{eqnarray}
\end{definition}
It follows from equation \eqref{Gamma} that
\begin{eqnarray}\label{orth}
h_{i j}\bullet E_k =0, && E_k\bullet \tilde h_{i j} =0.
\end{eqnarray}
\begin{remark}
Both the left and right second fundamental forms reduce to $h^0_{i
j} {\bf N}$ in the commutative limit, where $h^0_{i j}$ is the
standard second fundamental form and $ {\bf N}$ is the unit normal
vector.
\end{remark}

The Riemann curvature $R_{l k i j }= (\nabla_i \nabla_j- \nabla_j
\nabla_i) E_k\bullet E_l$ can be expressed in terms of the second
fundamental forms. Note that
\begin{eqnarray*}
R_{l k i j }&=&  \partial_j  E_k\bullet \partial_i E_l -
\partial_j E_k\bullet \tilde \nabla_i E_l - \partial_i E_k\bullet
\partial_j E_l +
\partial_i E_k\bullet \tilde \nabla_j E_l.
\end{eqnarray*}
By Definition \ref{sff},
\begin{eqnarray*}
R_{l k i j }&=& \partial_j E_k\bullet \tilde h_{i l} -
\partial_i E_k\bullet \tilde h_{j l}\\
&=& (\nabla_j E_k +h_{j k})\bullet \tilde h_{i l} - (\nabla_i E_k +
h_{i k})\bullet \tilde h_{j l}.
\end{eqnarray*}
Equation \eqref{orth} immediately leads to the following result.
\begin{lemma} The following generalized Gauss
equation holds:
\begin{eqnarray}
R_{l k i j } &=& h_{j k}\bullet \tilde h_{i l} - h_{i k}\bullet
\tilde h_{j l}.
\end{eqnarray}
\end{lemma}

Before closing this section, we mention that the Riemannian
structure of a noncommutative surface is a deformation of the
classical Riemannian structure of a surface by including quantum
corrections. The embedding into $\cA^3$ is not subject to any
constraints as the general theory stands. However, one may consider
particular noncommutative surfaces with embeddings satisfying extra
symmetry requirements similar to the way in which various star
products on $\R^3$ were obtained from the Moyal product on $\R^4$ in
Sections 4 and 5 in Ref. \cite{GLMV}.

\section{Examples}\label{examples}

In this section, we consider in some detail three concrete examples
of noncommutative surfaces: the noncommutative sphere, torus and
hyperboloid.

\subsection{Noncommutative sphere} \label{sphere}

Let $U=(0, \pi)\times(0, 2\pi)$, and we write $\theta$ and $\phi$
for $t_1$ and $t_2$ respectively. Let $X(\theta, \phi)=(X^1(\theta,
\phi), X^2(\theta, \phi), X^3(\theta, \phi))$ be given by
\begin{eqnarray}
X(\theta, \phi) &=& \left(\frac{\sin\theta\cos\phi}{\cosh\barh},
\frac{\sin\theta\sin\phi}{\cosh\barh}, \frac{\sqrt{\cosh 2\barh }
\cos\theta}{\cosh\barh}\right)
\end{eqnarray}
with the components being smooth functions in $(\theta, \phi)\in U$.
It can be shown that $X$ satisfies the following relation
\begin{eqnarray}\label{eq:sphere}
X^1\ast X^1 + X^2\ast X^2 + X^3\ast X^3 =1.
\end{eqnarray}
Thus we may regard the noncommutative surface defined by $X$ as an
analogue of the sphere $S^2$. We shall denote it by $S^2_\barh$ and
refer to it as a {\em noncommutative sphere}. We have
\begin{eqnarray*}
E_1 &=&\left(\frac{\cos\theta\cos\phi}{\cosh\barh},
\frac{\cos\theta\sin\phi}{\cosh\barh}, -\frac{\sqrt{\cosh 2\barh }
\sin\theta}{\cosh\barh}\right),\\
E_2&=&\left(-\frac{\sin\theta\sin\phi}{\cosh\barh},
\frac{\sin\theta\cos\phi}{\cosh\barh}, 0\right).
\end{eqnarray*}
The components  $g _{ij}=E _i \bullet E _j$  of the metric $g$ on
$S^2_\barh$ can now be calculated, and we obtain
\begin{eqnarray*}
\begin{aligned}
&g_{1 1} = 1 ,  \quad g_{2 2}= \sin^2\theta -
\frac{\sinh^2\barh}{\cosh^2\barh}\cos^2\theta  ,\\
&g_{1 2} =- g_{2 1}= \frac{\sinh\barh}{\cosh\barh}\left(\sin^2\theta
- \cos^2\theta\right).
\end{aligned}
\end{eqnarray*}

The components of this metric commute with one another as they
depend on $\theta$ only. Thus it makes sense to consider the usual
determinant $G$ of $g$. We have
\begin{eqnarray*}
\begin{aligned}
G =&\sin ^2 \theta  + \tanh ^2 \barh (\cos ^2 2 \theta -\cos
^2 \theta) \\
=&\sin ^2 \theta [1 + \tanh ^2 \barh (1-4\cos ^2  \theta)].
\end{aligned}
\end{eqnarray*}
The inverse metric is given by
\begin{eqnarray*}
\begin{aligned}
&g^{1 1} = \frac{\sin ^2 \theta -\tanh^2\barh \cos ^2 \theta}
{\sin ^2 \theta +\tanh^2\barh (\cos ^2 2\theta -\cos ^2 \theta)} ,  \\
&g^{2 2}= \frac{1}{\sin ^2 \theta +\tanh^2\barh (\cos ^2 2\theta -\cos ^2 \theta)} ,\\
&g^{1 2} =- g^{2 1}=\frac{\tanh \barh \cos 2 \theta} {\sin ^2 \theta
+\tanh^2\barh (\cos ^2 2\theta -\cos ^2 \theta)}.
\end{aligned}
\end{eqnarray*}

Now we determine the connection and curvature tensor of the
noncommutative sphere. The computations are quite lengthy, thus we
only record the results here. For the Christoffel symbols, we have
\begin{eqnarray*}
\begin{aligned}
& \Gamma _{1 1 1} = \tilde\Gamma _{1 1 1}=0, &\quad& \Gamma _{1 1 2}
= -\tilde\Gamma _{1 1 2} =
\sin 2\theta \tanh \barh,  \\
& \Gamma _{1 2 1} = -\tilde\Gamma _{1 2 1}=-\sin 2\theta \tanh
\barh,
&\quad& \Gamma _{1 2 2} = \tilde\Gamma _{1 2 2} =\frac{1}{2}\sin 2\theta (1+\tanh ^2 \barh),  \\
& \Gamma _{2 1 1} = -\tilde\Gamma _{2 1 1}=\sin 2\theta \tanh \barh,
&\quad& \Gamma _{2 1 2} = \tilde\Gamma _{2 1 2} =\frac{1}{2}\sin
2\theta (1+\tanh ^2 \barh),\\
&\Gamma _{2 2 1} = \tilde\Gamma _{2 2 1} =-\frac{1}{2}\sin 2\theta
(1+\tanh ^2 \barh), &\quad& \Gamma _{2 2 2} = -\tilde\Gamma _{2 2 2}
=\sin 2\theta \tanh \barh.
\end{aligned}
\end{eqnarray*}
Note that $\Gamma _{1 1 2}\ne \tilde\Gamma _{1 1 2}$ (c.f. Remark
\ref{Gammas}). We now find the asymptotic expansions of the
curvature tensors with respect to $\barh$:
\begin{eqnarray*}
\begin{aligned}
R _{1 1 1 2} = &2 \barh +(\frac{10}{3} +4 \cos 2 \theta)\barh ^3+O(\barh ^4),  \\
 R _{2 1 1 2} = &-\sin ^2 \theta -\frac{1}{2}(4+\cos 2 \theta -\cos 4
 \theta)
 \barh ^2+O(\barh
 ^4),\\
R _{1 2 1 2} = &\sin ^2 \theta +\frac{1}{2}(4+\cos 2 \theta -\cos 4
\theta) \barh ^2+O(\barh
 ^4),\\
R _{2 2 1 2} = &-2\sin ^2 \theta \barh -(\frac{5}{3}+\frac{4}{3}\cos
2 \theta -4 \cos 4 \theta) \barh ^3+O(\barh ^4).
\end{aligned}
\end{eqnarray*}
We can also compute asymptotic expansions of the Ricci curvature
tensor
\begin{eqnarray*}
\begin{aligned}
 R _{1 1} = &1+(6+4\cos 2\theta)\barh ^2 +O(\barh ^4),  \\
 R _{2 1} = &(2-\cos 2\theta)\barh +\frac{1}{3}
 (16+19\cos 2 \theta -6 \cos 4\theta )\barh ^3 +O(\barh ^4),  \\
 R _{1 2} = &(2+\cos 2\theta)\barh +\frac{1}{3}
 (16+29\cos 2 \theta +6 \cos 4\theta )\barh ^3 +O(\barh ^4),  \\
 R _{2 2} = &\sin ^2 \theta +\frac{1}{2}
 (3+5\cos 2 \theta -2\cos 4 \theta)\barh ^2+O(\barh ^4),
\end{aligned}
\end{eqnarray*}
and the scalar curvature
\begin{eqnarray*}
 R  = 2+4(3+4\cos 2 \theta) \barh ^2+O(\barh
 ^4).
\end{eqnarray*}

By setting $\barh=0$, we obtain from the various curvatures of
$S^2_\barh$ the corresponding objects for the usual sphere $S^2$.
This is a useful check that our computations above are accurate.

\subsection{Noncommutative torus} \label{torus}

This time we shall take $U=(0, 2\pi)\times(0, 2\pi)$, and denote a
point in $U$ by $(\theta, \phi)$. Let $X(\theta, \phi)=(X^1(\theta,
\phi), X^2(\theta, \phi), X^3(\theta, \phi))$ be given by
\begin{eqnarray}
X(\theta, \phi) &=& \left((a+\sin\theta)\cos\phi,
(a+\sin\theta)\sin\phi,  \cos\theta\right)
\end{eqnarray}
where $a>1$ is a constant. Classically $X$ is the torus. When we
extend scalars from $\R$ to $\R[[\barh]]$ and impose the star
product on the algebra of smooth functions, $X$ gives rise to a
noncommutative torus, which will be denoted by $T ^2 _\barh$. We
have
\begin{eqnarray*}
E_1 &=&\left(\cos\theta\cos\phi, \cos\theta\sin\phi,
-\sin\theta\right),\\
E_2&=&\left(-(a+\sin\theta)\sin\phi, (a+\sin\theta)\cos\phi,
0\right).
\end{eqnarray*}
The components  $g _{ij}=E _i \bullet E _j$  of the metric $g$ on
$T^2_\barh$ take the form
\begin{eqnarray*}
\begin{aligned}
&g_{1 1} = 1 +\sinh ^2 \barh \cos 2\theta,  \\
&g_{2 2}= (a+\cosh\barh \sin\theta) ^2 -\sinh^2\barh \cos^2\theta  ,\\
&g_{1 2} =- g_{2 1}=- \sinh\barh \cosh\barh \cos 2\theta +a \sinh
\barh \sin \theta.
\end{aligned}
\end{eqnarray*}
As they depend only on $\theta$, the components of the metric
commute with one another. The inverse metric is given by
\begin{eqnarray*}
\begin{aligned}
&g^{1 1} = \frac{(a+\cosh\barh \sin\theta) ^2 -\sinh^2\barh
\cos^2\theta}{G},  \\
&g^{2 2}= \frac{1+\sinh ^2 \barh \cos 2\theta}{G} ,\\
&g^{1 2} =- g^{2 1}=\frac{\sinh\barh \cosh\barh \cos 2\theta +a
\sinh \barh \sin \theta}{G},
\end{aligned}
\end{eqnarray*}
where $G$ is the usual determinant of $g$ given by
\begin{eqnarray*}
G =(\sin \theta  + a\cosh \barh ) ^2-a ^2 \sin ^2 \theta \sinh ^2
\barh.
\end{eqnarray*}

Now we determine the curvature tensor of the noncommutative torus.
The computations can be carried out in much the same way as in the
case of the noncommutative sphere, and we merely record the results
here. For the connection, we have
\begin{eqnarray*}
\begin{aligned}
& \Gamma _{1 1 1} = - \sin 2\theta \sinh ^2 \barh, &\quad& \Gamma
_{1 1 2} = a \cos \theta \sinh \barh + \sin 2\theta \sinh \barh\cosh
\barh,  \\
& \Gamma _{1 2 1} = -\sin 2\theta  \sinh \barh \cosh \barh,
&\quad& \Gamma _{1 2 2} = a \cos \theta \cosh \barh+\frac{1}{2}\sin 2\theta \cosh 2\barh,  \\
& \Gamma _{2 1 1} = -\sin 2\theta \sinh \barh \cosh \barh, &\quad&
\Gamma _{2 1 2} = a \cos \theta \cosh \barh +\frac{1}{2}\sin
2\theta \cosh 2 \barh,\\
&\Gamma _{2 2 1} = -a \cos \theta \cosh \barh -\frac{1}{2}\sin
2\theta \cosh \barh, &\quad& \Gamma _{2 2 2} = 2a \cos\theta \sinh
\barh + \sin 2\theta \sinh \barh\cosh \barh.
\end{aligned}
\end{eqnarray*}
We can find the asymptotic expansions of the curvature tensors with
respect to $\barh$:
\begin{eqnarray*}
\begin{aligned}
 R _{1 1 1 2} = &\frac{2\sin \theta (1+a \sin \theta)}{a+\sin \theta} \barh +O(\barh ^3),  \\
 R _{2 1 1 2} = &-\sin \theta (a+\sin \theta)+O(\barh ^2),\\
 R _{1 2 1 2} = &\sin \theta (a+\sin \theta)+O(\barh ^2),\\
 R _{2 2 1 2} = &-2\sin ^2 \theta (1+a \sin \theta) \barh +O(\barh ^3).
\end{aligned}
\end{eqnarray*}
We can also compute asymptotic expansions of the Ricci curvature
tensor
\begin{eqnarray*}
\begin{aligned}
 R _{1 1} = &\frac{\sin \theta}{a+\sin \theta} +O(\barh ^2),  \\
 R _{2 1} = &-\frac{\sin \theta(-3a +5a \cos \theta -(5+2a ^2)
 \sin \theta +\sin 3\theta)}{2(a+\sin \theta) ^2}\barh +O(\barh ^3),  \\
 R _{1 2} = &\frac{\sin \theta(a+\cos 2 \theta +a \sin \theta)}{a+\sin \theta}\barh +O(\barh ^3),  \\
 R _{2 2} = &\sin \theta (a+\sin \theta)+O(\barh ^2),
\end{aligned}
\end{eqnarray*}
and the scalar curvature
\begin{eqnarray*}
R  = \frac{2\sin \theta}{a+\sin \theta}+O(\barh^2).
\end{eqnarray*}
By setting $\barh=0$, we obtain from the various curvatures of
$T^2_\barh$ the corresponding objects for the usual torus $T^2$.

\subsection{Noncommutative hyperboloid} \label{hyperbolid}

Another simple example is the noncommutative analogue of the
hyperboloid described by $X= (x, y, {\sqrt{1+x^2+y^2}})$. One may
also change the parametrization and consider instead
\begin{eqnarray}
X(r, \phi) &=& \left(\sinh r \cos \phi, \sinh r \sin \phi, \cosh
r\right)
\end{eqnarray}
on $U=(0, \infty)\times(0, 2\pi)$, where a point in $U$ is denoted
by $(r, \phi)$. When we extend scalars from $\R$ to $\R[[\barh]]$
and impose the star product on the algebra of smooth functions
(defined by \eqref{product} with $t_1=r$ and $t_2=\phi$), $X$ gives
rise to a noncommutative hyperboloid, which will be denoted by $H ^2
_\barh$. We have
\begin{eqnarray*}
E_1 &=&\left(\cosh r \cos\phi, \cosh r \sin\phi, \sinh r\right),\\
E_2&=&\left(-\sinh r \sin\phi, \sinh r \cos\phi, 0\right).
\end{eqnarray*}
The components  $g _{ij}=E _i \bullet E _j$  of the metric $g$ on
$H^2_\barh$ take the form
\begin{eqnarray*}
\begin{aligned}
&g_{1 1} = \cos ^2 \barh \cosh 2r,  \\
&g_{2 2}= \frac{1}{2} \left(-1+\cos 2\barh \cosh 2r\right),\\
&g_{1 2} =- g_{2 1}=- \frac{1}{2}\sin 2\barh \cosh 2r.
\end{aligned}
\end{eqnarray*}
As they depend only on $r$, the components of the metric commute
with one another. The inverse metric is given by
\begin{eqnarray*}
\begin{aligned}
&g^{1 1} =\frac{\sec^2 \barh}{2\sinh^2 r }\left(\cos 2\barh-\frac{1}{\cosh 2r} \right),\\
&g^{2 2}= \frac{1}{\sinh^2 r},\\
&g^{1 2} =- g^{2 1}=\frac{\tan\barh}{\sinh^2 r}.
\end{aligned}
\end{eqnarray*}

Now we determine the curvature tensor of the noncommutative
hyperboloid. For the connection, we have
\begin{eqnarray*}
\begin{aligned}
& \Gamma _{1 1 1} = \cos^2 \barh \sinh 2r, &\quad& \Gamma _{1 1 2} =
-\frac{1}{2}\sin 2\barh \sinh 2r,\\
& \Gamma _{1 2 1} = \frac{1}{2}\sin 2\barh \sinh 2r, &\quad& \Gamma
_{1 2 2} = \frac{1}{2}\cos 2\barh \sinh 2r,  \\
& \Gamma _{2 1 1} =\frac{1}{2}\sin 2\barh \sinh 2r, &\quad& \Gamma
_{2 1 2} = \frac{1}{2}\cos 2\barh \sinh 2r,\\
&\Gamma _{2 2 1} = -\frac{1}{2}\cos 2\barh \sinh 2r, &\quad& \Gamma
_{2 2 2} =\frac{1}{2}\sin 2\barh \sinh 2r.
\end{aligned}
\end{eqnarray*}
We can find the asymptotic expansions of the curvature tensors with
respect to $\barh$:
\begin{eqnarray*}
\begin{aligned}
 R _{1 1 1 2} = &\frac{2}{\cosh 2r}\barh +O(\barh ^2),\\
 R _{2 1 1 2} = &-\frac{\sinh ^2 r}{\cosh 2r} +O(\barh ^2),\\
 R _{1 2 1 2} = &\frac{\sinh ^2 r}{\cosh 2r} +O(\barh ^2),\\
 R _{2 2 1 2} = &-\frac{\cosh 2r +\sinh ^2 2r}{\cosh 2r}\barh +O(\barh ^3).
\end{aligned}
\end{eqnarray*}
We can also compute asymptotic expansions of the Ricci curvature
tensor
\begin{eqnarray*}
\begin{aligned}
 R _{1 1} = &\frac{1}{\cosh 2r} +O(\barh ^2),  \\
 R _{2 1} = &\frac{\coth^2 r (2\cosh 2r -1)}{\cosh 2r} \barh +O(\barh ^3),  \\
 R _{1 2} = &\frac{\cosh 2r +2}{\cosh^2 2r}\barh +O(\barh ^3),  \\
 R _{2 2} = &\frac{\sinh ^2 r}{\cosh ^2 2r}+O(\barh ^2),
\end{aligned}
\end{eqnarray*}
and the scalar curvature
\begin{eqnarray*}
R  = \frac{2}{\cosh ^2 2r}+O(\barh^2).
\end{eqnarray*}
By setting $\barh=0$, we obtain from the various curvatures of
$H^2_\barh$ the corresponding objects for the usual hyperboloid
$H^2$.

\section{Noncommutative $n$-dimensional surfaces}\label{general}

One can readily generalize the theory of Section \ref{surfaces} to
higher dimensions, and we shall do this here. Noncommutative
Bianchi identities will also be obtained.

Again for the sake of explicitness we restrict attention to the
Moyal product on the smooth functions. However, as we shall see in
Section \ref{transformations}, it will be necessary to consider more
general star-products in order to discuss ``general coordinate
transformations" of noncommutative surfaces.

\subsection{Noncommutative $n$-dimensional surfaces}
We take a region $U$ in $\R^n$ for a fixed $n$, and write the
coordinate of $t\in U$ as $(t_1, t_2, \dots, t_n)$. Let $\cA$ denote
the set of the smooth functions on $U$ taking values in
$\R[[\barh]]$. Fix any constant skew symmetric $n\times n$ matrix
$\theta$. The Moyal product on $\cA$ is defined by the following
generalization of equation \eqref{product}:
\begin{eqnarray}\label{multiplication}
f\ast g = \lim_{t'\rightarrow t}\ \exp{\left(\barh \sum_{i j}
\theta_{i j}\partial_i\partial_j^\prime\right)}f(t) g(t')
\end{eqnarray}
for any $f, g\in \cA$. Such a multiplication is known to be
associative. Since $\theta$ is a constant matrix, the Leibniz rule
\eqref{Leibniz} remains valid in the present case:
\[ \partial_i(f\ast g)= \partial_if\ast g + f\ast \partial_i g. \]

For any fixed positive integer $m$, we can define a dot-product
\begin{eqnarray}\label{dotproduct-dimm}
 \bullet: \cA^m\otimes_{\R[[\barh]]} \cA^m\longrightarrow
\cA^m \end{eqnarray}
 by generalizing \eqref{dotproduct} to
$A\bullet B$ $=$ $a^i\ast b_i$ for all $A=(a^1, \dots, a^m)$ and
$B=(b_1, \dots, b_m)$ in $\cA^m$. As before, the dot-product is a
map of two-sided $\cA$-modules.

Assume $m>n$. For $X\in \cA^m$, we let $E_i=\partial_i X$, and
define $g_{i j}=E_i \bullet E_j$. Denote by $g=(g_{i j})$ the
$n\times n$ matrix with entries $g_{i j}$.
\begin{definition}\label{manifold}
If $g \mod \barh$ is invertible over $U$, we shall call $X$ a {\em
noncommutative $n$-dimensional surface} embedded in $\cA^m$, and
call $g$ the metric of $X$.
\end{definition}

The discussion on the metric in Section \ref{surfaces} carries over
to the present situation; in particular, the invertibility of $g
\mod \barh$ implies that there exists a unique inverse $(g^{i j})$.
Now as in Section \ref{surfaces}, we define the left tangent bundle
$TX$ (respectively right tangent bundle $\tilde TX$) of the
noncommutative surface as the left (respectively right)
$\cA$-submodule of $\cA^m$ generated by the elements $E_i$. The fact
that the metric $g$ belongs to $GL_n(\cA)$ enables us to show that
the left and right tangent bundles are projective $\cA$-modules.

The connection $\nabla_i$ on the left tangent bundle will be defined
in the same way as in Section \ref{surfaces}, namely, by the
composition of the derivative $\partial_i$ with the projection of
$\cA^m$ onto the left tangent bundle. The connection
$\tilde\nabla_i$ on the right tangent bundle is defined similarly.
Then $\nabla_i$ and $\tilde\nabla_i$ satisfy the analogous equation
\eqref{connection}, and are compatible with the metric in the same
sense as Proposition \ref{prop:compatible}.

One can show that
\[
{[\nabla_i , \nabla_j]}: TX\longrightarrow TX, \quad
{[\tilde\nabla_i , \tilde\nabla_j]}: \tilde TX\longrightarrow \tilde
TX
\]
are left and right $\cA$-module homomorphisms respectively. This
allows us to define Riemann curvatures of the tangent bundles as in
equation \eqref{curvature}. Then the formulae given in Lemma
\ref{curvature1} are still valid when the indices in the formulae
are assumed to take values in $\{1, 2, \dots, n\}$. Furthermore, the
left and right Riemann curvatures remain equal in the sense of
Proposition \ref{curvature3}.

\begin{remark}\label{pseudo}
One may define a dot-product  $\bullet: \cA^m\otimes_{\R[[\barh]]}
\cA^m\longrightarrow \cA^m$ with a Minkowski signature by
\[
A\bullet B = a_0\ast b_0 - \sum_{i=1}^{m-1} a_i\ast b_i
\]
for any $A=(a_0, a_1, \dots, a_{m-1})$ and $B$ $=$ $(b_0, b_1,
\dots, b_{m-1})$ in $\cA^m$. This is still a map of two-sided
$\cA$-modules. Then the afore developed theory can be adapted to
this setting, leading to a theory of a noncommutative surface
embedded in $\cA^m$ with a Minkowski signature.
\end{remark}

For the sake of being concrete, we shall consider only
noncommutative surfaces with Euclidean signature hereafter.

\subsection{Bianchi identities}\label{Bianchi}
We examine properties of the Riemann curvature for arbitrary $n$ and
$m$. The main result in this subsection is the noncommutative
analogues of Bianchi identities.

Define $E^i$ and $\tilde E^l$ as in \eqref{E-upper}. Then
\begin{eqnarray}\label{upper-gamma}
\nabla_p E^l = -\tilde\Gamma_{p k}^l \ast E^k, \quad
\tilde\nabla_p \tilde E^l = -\tilde E^k \ast\Gamma_{p k}^l.
\end{eqnarray}
These relations will be needed presently. Let
\begin{eqnarray}
R^l_{k i j; p} = \partial_p R^l_{k i j} - \Gamma_{p k}^r\ast
R^l_{r i j}- \Gamma_{p i}^r\ast R^l_{r j k} - \Gamma^r_{p j}\ast
R^l_{r k i} + R^r_{k i j} \ast\Gamma_{r p}^l.
\end{eqnarray}

\begin{theorem}\label{BIs}
The Riemann curvature $R^i_{j k l}$ satisfies the following
relations
\begin{eqnarray}
R^l_{i j k} + R^l_{ j k i} + R^l_{k i j}=0,&\qquad & R^l_{k i j;p}
+ R^l_{k j p; i} + R^l_{k p i; j}=0,
\end{eqnarray}
which will be referred to as the first and second noncommutative
Bianchi identities respectively.
\end{theorem}
\begin{proof}
It follows from the relation $\nabla_i E_j = \nabla_j E_i$ that
\[
[\nabla_i, \nabla_j] E_k + [\nabla_j, \nabla_k] E_i+[\nabla_k,
\nabla_i] E_j = 0.
\]
This immediately leads to
\[
g([\nabla_i, \nabla_j] E_k, \tilde E^l)  + g([\nabla_j, \nabla_k]
E_i, \tilde E^l)+g([\nabla_k, \nabla_i] E_j, \tilde E^l) = 0.
\]
Using the definition of the Riemann curvature in this relation, we
obtain the first Bianchi identity.

To prove the second Bianchi identity, note that
\[
-\partial_p R^l_{k i j} + g(\nabla_p [\nabla_i, \nabla_j] E_k,
\tilde E^l) + g([\nabla_i, \nabla_j] E_k, \tilde \nabla_p \tilde
E^l)=0.
\]
Cyclic permutations of the indices $p, i, j$ lead to two further
relations. Adding all the three relations together, we arrive at
\begin{eqnarray}
\begin{aligned}\label{intermediate}
& & -\partial_p R^l_{k i j} + g([\nabla_i, \nabla_j] \nabla_p E_k,
\tilde E^l) + g([\nabla_i, \nabla_j] E_k, \tilde \nabla_p \tilde
E^l)&&\\
& & - \partial_i R^l_{k j p} + g([\nabla_j, \nabla_p] \nabla_i
E_k, \tilde E^l) + g([\nabla_j, \nabla_p] E_k, \tilde \nabla_i
\tilde
E^l)&&\\
& & - \partial_j R^l_{k p i} + g([\nabla_p,  \nabla_i] \nabla_j
E_k, \tilde E^l) + g([\nabla_p,  \nabla_i] E_k,  \tilde \nabla_j
\tilde E^l)&=&0,
\end{aligned}
\end{eqnarray}
where we have used the following variant of the Jacobian identity
\[
\nabla_p [\nabla_i, \nabla_j] + \nabla_i [\nabla_j, \nabla_p] +
\nabla_j [\nabla_p, \nabla_i] =  [\nabla_i, \nabla_j]\nabla_p +
[\nabla_j, \nabla_p]\nabla_i  + [\nabla_p, \nabla_i]\nabla_j.
 \]
By a tedious calculation one can show that
\begin{eqnarray*}
{\mathcal Q}_{i j k p}&:=& [\nabla_j, \nabla_k]\nabla_p E_i +
[\nabla_k, \nabla_i]\nabla_p E_j \\
&&+ [\nabla_p, \nabla_k]\nabla_i E_j
+ [\nabla_k, \nabla_j]\nabla_i E_p\\
&&+ [\nabla_i, \nabla_k]\nabla_j E_p + [\nabla_k, \nabla_p]\nabla_j
E_i
\end{eqnarray*}
is identically zero. Now we add $g({\mathcal Q}_{i j k p}, \tilde
E^l)$ to the left-hand side of \eqref{intermediate}, obtaining an
identity with fifteen terms on the left. Then the second Bianchi
identity can be read off this equation by recalling
\eqref{upper-gamma}.
\end{proof}

\subsection{Einstein's equation}\label{sect:Einstein}
Recall that in classical Riemannian geometry, the second Bianchi
identity suggests the correct form of Einstein's equation. Let us
make some preliminary analysis of this point here.  As we lack
guiding principles for constructing an analog of Einstein's
equation, the material of this subsection is of a rather speculative
nature.

In Section \ref{surfaces},  we introduced the Ricci curvature $R_{i
j}$ and scalar curvature $R$. Their definitions can be generalized
to higher dimensions in an obvious way. Let
\begin{eqnarray}
R^i_j = g^{i k}\ast R_{k j},
\end{eqnarray}
then the scalar curvature is $R=R^i_i$. Let us also introduce the
following object:
\begin{eqnarray}\label{Theta}
\Theta^l_p :=g([\nabla_p, \nabla_i]E^i,\,  \tilde E^l)= g^{i
k}\ast R^l_{k p i}.
\end{eqnarray}
In the commutative case, $\Theta^l_p$ coincides with $R^l_p$, but
it is no longer true in the present setting. However, note that
\begin{eqnarray}\label{Scalar2}
\Theta^l_l =g ^{ik} \ast R ^l _{kli} = g ^{ik} \ast R _{ki} =R.
\end{eqnarray}

By first contracting the indices $j$ and $l$ in the second Bianchi
identity, then raising the index $k$ to $i$ by multiplying the
resulting identity by $g^{i k}$ from the left and summing over $i$,
we obtain the identity
\begin{eqnarray*}
\begin{aligned}
0&=\partial_p R & - &\partial_i R^i_p&+& g\left([\nabla_i,
\nabla_l]\nabla_p E^i, \tilde E^l \right)&+&g\left([\nabla_l,
\nabla_p]\nabla_i E^i, \tilde E^l \right)\\
& &-&\partial_l \Theta^l_p&+& g\left([\nabla_i, \nabla_l]E^i,
\tilde \nabla_p \tilde E^l \right)&+& g\left([\nabla_p,
\nabla_i]E^i, \tilde \nabla_l \tilde E^l \right)\\
& && &+&g\left([\nabla_p, \nabla_i]\nabla_l E^i, \tilde E^l
\right)&+& g\left([\nabla_l, \nabla_p]E^i, \tilde \nabla_i \tilde
E^l \right).
\end{aligned}
\end{eqnarray*}
Let us denote the sum of the last two terms on the right-hand side
by $\varpi_p$. Then
\begin{eqnarray*}
\varpi_p&=& g^{i k}\ast R^r_{k p l}\ast \Gamma^l_{r
i}-\tilde\Gamma^i_{l r}\ast g^{r k}\ast R^l_{k p i}.
\end{eqnarray*}
In the commutative case, $\varpi_p$ vanishes identically for all
$p$. However in the noncommutative setting, there is no reason to
expect this to happen. Let us now define
\begin{eqnarray}
\begin{aligned}
R^i_{p;i} &=& \partial_i R^i_p - \tilde\Gamma^i_{p r}\ast R^r_i
+ \tilde\Gamma^i_{i r}\ast R^r_p, \\
\Theta^l_{p;l} &=& \partial_l \Theta^l_p -
\Theta^r_l\ast\Gamma^l_{r p}+ \Theta^r_p\ast\Gamma^l_{r
l}&-\varpi_p.
\end{aligned}
\end{eqnarray}
Then the second Bianchi identity implies
\begin{eqnarray}\label{divergence}
R^i_{p;i} + \Theta^i_{p;i} - \partial_p R =0.
\end{eqnarray}

The above discussions suggest that Einstein's equation no longer
takes its usual form in the noncommutative setting, but we have not
been able to formulate a basic principle which enables us to {\em
derive} a noncommutative analogue of Einstein's equation. However,
formulae \eqref{divergence} and \eqref{Scalar2} seem to suggest that
the following is a reasonable proposal for a noncommutative Einstein
equation in the vacuum:
\begin{eqnarray} \label{Einstein}
R^i_j + \Theta^i_j- \delta^i_j R &=&0.
\end{eqnarray}
We were informed by J. Madore that in other contexts of
noncommutative general relativity, it also appeared to be necessary
to include an object analogous to $\Theta_j^i$ in the Einstein
equation.

\section{General coordinate transformations}\label{transformations}

We investigate the effect of ``general coordinate transformations"
on noncommutative $n$-dimensional surfaces. This requires us to
consider noncommutative surfaces defined over $\cA$ endowed with
star-products more general than the Moyal product. This should be
compared with Refs. \cite{C01, C04, ADMW1, ADMW2}, where the only
``general coordinate transformations" allowed were those keeping the
Moyal product intact.

For the sake of being concrete, we assume that the noncommutative
surface has Euclidean signature.

\subsection{Gauge transformations}\label{transformations-sub}
Denote by $\cG(\cA)$ the set of $\R[[\barh]]$-linear maps $\phi:
\cA\longrightarrow \cA$ satisfying the following conditions
\begin{eqnarray}
\phi(1)=1, \quad \phi= \exp\left(\sum_i \epsilon^i\partial_i\right)
\ \text{mod}\ \barh,
\end{eqnarray}
where $\epsilon^i$ are smooth functions on $U$. Then clearly we
have the following result.
\begin{lemma} The set $\cG(\cA)$ forms a subgroup of the
automorphism group of $\cA$ as $\R[[\barh]]$-module.
\end{lemma}

For any given $\phi\in\cG(\cA)$, define an $\R[[\barh]]$-linear map
\begin{eqnarray}
\ast_\phi: \cA\otimes_{\R[[\barh]]}\cA\longrightarrow \cA, \quad
 f\otimes g \mapsto f\ast_\phi g :=
\phi^{-1}\left( \phi(f) \ast \phi(g)\right).
\end{eqnarray}
\begin{lemma}\label{gauge-equiv}
\begin{enumerate}
\item\label{gauge-equiv:a} The map $\ast_\phi$ is associative, thus there exists the
associative algebra $(\cA, \ast_\phi)$ over $\R[[\barh]]$.
Furthermore, $\phi: (\cA, \ast_\phi) \longrightarrow (\cA, \ast)$ is
an algebra isomorphism.

\item\label{gauge-equiv:b} Let
$\ast^{\phi} = \ast_{\phi^{-1}}$, then for any $\phi, \psi\in
\cG(\cA)$
\begin{eqnarray}\label{group-structure}
\psi\left(\psi^{-1}(f)\ast^\phi\psi^{-1}(g)\right) =
f\ast^{\psi\phi} g.
\end{eqnarray}
In this sense the definition of the new star-products respects the
group structure of $\cG(\cA)$.
\end{enumerate}
\end{lemma}
\begin{proof}
Because of the importance of this lemma for later discussions,  we
sketch a proof for it here, even though one can easily deduce a
proof from Ref. \cite{Ge}.

For $f, g, h\in \cA$, we have
\begin{eqnarray*}
(f\ast_\phi g)\ast_\phi h &=&  \phi^{-1}\left((\phi(f)\ast\phi(
g))\ast \phi(h) \right)\\
&=&\phi^{-1}\left(\phi(f)\ast(\phi(
g)\ast\phi(h)) \right)\\
&=&\phi^{-1}\left(\phi(f)\ast\phi(
g\ast_\phi h) \right)\\
&=& f \ast_\phi (g\ast_\phi h),
\end{eqnarray*}
which proves the associativity of the new star-product. As $\phi$ is
an $\R[[\barh]]$-module isomorphism by definition, we only need to
show that it preserves multiplications in order to establish the
isomorphism between the algebras. Now $\phi(f\ast_\phi g ) = \phi(f)
\ast \phi(g)$. This proves part (\ref{gauge-equiv:a}).

Part (\ref{gauge-equiv:b}) can be proven by unraveling the left-hand
side of \eqref{group-structure}.
\end{proof}

Adopting the terminology of Drinfeld from the context of quantum
groups, we call an automorphism $\phi \in \cG(\cA)$ a {\em gauge
transformation}, and call $\cG(\cA)$ the {\em gauge group}. The star
product $\ast_\phi$ will be said to be {\em gauge equivalent} to the
Moyal product \eqref{multiplication}. However, note that our notion
of gauge transformations is slightly more general than that in
deformation theory \cite{Ge}, where the only type of gauge
transformations allowed are of the special form
\[
\phi=id  + \barh \phi_1 + \barh^2 \phi_2 + \dots,
\]
with $\phi_i$ being $\R$-linear maps on the space of smooth
functions on $U$ such that $\phi_i(1)=0$ for all $i$. Such gauge
transformations form a subgroup of $\cG(\cA)$.

\begin{remark}
The prime aim of the deformation theory \cite{Ge} is to classify
the gauge equivalence classes of deformations in this restricted
sense but for arbitrary associative algebras. The seminal paper
\cite{Ko} of Kontsevich provided an explicit formula for a
star-product from each gauge equivalence class of deformations of
the algebra of functions on a Poisson manifold.
\end{remark}

\begin{remark}
General star-products gauge equivalent to the Moyal product were
evaluated explicitly up to the third order in $\barh$ in Ref.
\cite{Z}. In Ref. \cite{GGR}, position-dependent
star-products were also investigated and the ultra-violet
divergences of a quantum $\phi^4$ theory on $4$-dimensional spaces
with such products were analyzed.
\end{remark}

Given an element $\phi$ in the group $\cG(\cA)$, we denote
 \[ u^i:=\phi^{-1}(t^i), \quad i=1, 2, \dots, n,\]
and refer to $t\mapsto u$ as a {\em general coordinate
transformation} of $U$. Define $\R[[\barh]]$-linear operators on
$\cA$ by
\begin{eqnarray}\label{dphi}
\partial_i^\phi = \phi^{-1}\circ \partial_i\circ \phi.
\end{eqnarray}
\begin{lemma}
The operators  $\partial_i^\phi$ have the following properties
\[
\partial_i^\phi \circ \partial_j^\phi - \partial_j^\phi \circ
\partial_i^\phi=0, \qquad
\partial_i^\phi \phi^{-1}(t^j) = \delta_i^j,
\]
and also satisfy the Leibniz rule
\[
\partial_i^\phi(f\ast_\phi g) = \partial_i^\phi(f) \ast_\phi g + f\ast_\phi
\partial_i^\phi(g), \quad \forall f, g\in \cA.
\]
\end{lemma}
\begin{proof} The proof is  easy but very illuminating. We have
\[
\partial_i^\phi \circ \partial_j^\phi - \partial_j^\phi \circ
\partial_i^\phi = \phi^{-1}\circ(\partial_i \partial_j -\partial_j
\partial_i)\circ \phi=0.
\]
Also, $\partial_i^\phi \phi^{-1}(t^j)= \phi^{-1}(\partial_i
t^j)=\delta_i^j$, since $\phi$ maps a constant function to itself.

To prove the Leibniz rule, we note that
\begin{eqnarray*}
\partial_i^\phi(f\ast_\phi g) &=& \phi^{-1} \left(\partial_i
(\phi(f)\ast\phi(g)) \right)\\
&=&\phi^{-1} (\partial_i \phi(f)\ast\phi(g)) +  \phi^{-1} (
\phi(f)\ast\partial_i \phi(g)) \\
&=&\phi^{-1} \left(\phi(\partial_i^\phi f)\ast\phi(g)\right) +
\phi^{-1}
\left(\phi(f)\ast\phi(\partial_i^\phi g)\right) \\
&=&\partial_i^\phi f\ast_\phi g + f\ast_\phi\partial_i^\phi g.
\end{eqnarray*}
This completes the proof of the lemma.
\end{proof}
The Leibniz rule plays a crucial role in constructing noncommutative
surfaces over $(\cA, \ast_\phi)$.

\subsection{Reparametrizations of noncommutative surfaces}\label{reparametrizations}

The construction of noncommutative surfaces over $(\cA, \ast)$ works
equally well over $(\cA, \ast_\phi)$ for any $\phi \in \cG(\cA)$.
Regard $\cA^m\otimes_{\R[[\barh]]}\cA^m$ as a two-sided $(\cA,
\ast_\phi)$-module, and define the new dot-product
\[\bullet_\phi:  \cA^m\otimes_{\R[[\barh]]}\cA^m \longrightarrow \cA  \]
by $A\bullet_\phi B=a^i\ast_\phi b_i$ for any $A=(a^1, a^2, \dots,
a^m)$ and $B=(b_1, b_2, \dots, b_m)$ in $\cA^m$. It is obviously a
map of two-sided $(\cA, \ast_\phi)$-modules. For an element
\[ X^\phi= (X^1, X^2, \dots, X^m) \]
in the free two-sided $(\cA, \ast_\phi)$-module $\cA^m$, we define
\[
E_i^\phi = \partial_i^\phi X^\phi,\quad \gphi_{i j}=
E_i^\phi\bullet_\phi E_j^\phi,
\]
where $\phi$ acts on $\cA^m$ in a componentwise way. As in Section
\ref{surfaces}, let
\[
\gphi= \left(\gphi_{i j}\right)_{i, j=1, \dots, n}.
\]
We shall say that $X$ is an $n$ dimensional noncommutative surface
with metric $\gphi$ if $\gphi\mod\barh$ is invertible. In this case,
$\gphi$ has an inverse $\left(\gphi^{i j}\right)$.

The left tangent bundle $TX^\phi$ and right tangent bundle $\tilde
TX^\phi$ of $X^\phi$ are now respectively the left and right
$(\cA, \ast_\phi)$-modules generated by $E_i^\phi$, $i=1, 2,
\dots, n$.  The metric $\gphi$ leads to a two-sided $(\cA,
\ast_\phi)$-module map
\[
\gphi: TX^\phi\otimes_{\R[[\barh]]}\tilde TX^\phi\longrightarrow
\cA, \quad Z\otimes W \mapsto Z\bullet_\phi W,
\]
which is the restriction  of the dot-product $\bullet_\phi$ to
$TX^\phi\otimes_{\R[[\barh]]}\tilde TX^\phi$. By using this map, we
can decompose $\cA^m$ into
\begin{eqnarray*}
\cA^m = TX^\phi \oplus (TX^\phi)^\bot, &\qquad& \text{as left
$(\cA, \ast_\phi)$-module}, \\
\cA^m = \tilde TX^\phi \oplus (\tilde TX^\phi)^\bot, &\qquad&
\text{as right $(\cA, \ast_\phi)$-module},
\end{eqnarray*}
where $(TX^\phi)^\bot$ is orthogonal to $\tilde TX^\phi$ and
$(\tilde TX^\phi)^\bot$ is orthogonal to $TX^\phi$ with respect to
the map induced by the metric.

As in Definition \ref{nabla}, the operators
\[
\nabla_i^\phi:
TX^\phi\longrightarrow TX^\phi, \qquad \tilde\nabla_i^\phi: \tilde
TX^\phi\longrightarrow \tilde TX^\phi
\]
are defined to be the compositions of $\partial_i^\phi$ with the
projections of $\cA^m$ onto the left and right tangent bundles
respectively. Thus, for any $Z\in TX^\phi$ and $\ W\in \tilde
TX^\phi$,
\begin{eqnarray}\label{nablaphi}
\gphi(\nabla_i^\phi Z, W)=\partial_i^\phi Z\bullet_\phi W, \quad
\gphi(Z, \tilde\nabla_i^\phi W)= Z\bullet_\phi
\partial_i^\phi W.
\end{eqnarray}

By using the Leibniz rule for $\partial_i^\phi$, we can show that
the analogous equations of \eqref{linear} are satisfied by
$\nabla_i^\phi$ and $\tilde\nabla_i^\phi$, namely, for all $Z\in
TX^\phi$, $\ W\in \tilde TX^\phi$ and $f\in\cA$,
\begin{eqnarray}\label{connection}
\begin{aligned}
\nabla_i^\phi (f\ast_\phi Z) &=& \partial^\phi_i f\ast_\phi Z +
f\ast_\phi \nabla_i^\phi Z, \\
\tilde\nabla_i^\phi (W\ast_\phi f) &=& W\ast_\phi\partial^\phi_i f
+ \tilde\nabla_i^\phi W\ast_\phi f.
\end{aligned}
\end{eqnarray}
Furthermore, the operators are metric compatible:
\[
\partial^\phi_i  \gphi(Z, \ W)=\gphi(\nabla_i^\phi Z, \ W)
+ \gphi(Z, \ \tilde\nabla_i^\phi W),
\quad \forall Z\in TX^\phi, \ W\in \tilde TX^\phi.
\]
Thus the two sets $\{\nabla_i^\phi\}$ and $\{\tilde\nabla_i^\phi\}$
define connections on the left and right tangent bundles
respectively.

The Christoffel symbols  ${\, }^\phi\Gamma_{i j}^k$ and ${\,
}^\phi\tilde \Gamma_{i j}^k$ in the present context are also
defined in the same way as before:
\begin{eqnarray*}
\nabla_i^\phi E_j^\phi= {\, }^\phi\Gamma_{i j}^k \ast_\phi
E_k^\phi, && \tilde \nabla_i^\phi E_j^\phi= E_k^\phi \ast_\phi {\,
}^\phi\tilde \Gamma_{i j}^k.
\end{eqnarray*}
Then we have
\[
{\, }^\phi\Gamma_{i j}^k = \partial_i^\phi E_j^\phi\bullet_\phi
E_l^\phi\ast_\phi \gphi^{l k}, \qquad {\, }^\phi\tilde\Gamma_{i
j}^k=\gphi^{k l}\ast_\phi E_l^\phi\bullet_\phi
\partial_i^\phi E_j^\phi.
\]
These formulae are of the same form as those in equation
\eqref{Gamma1}.

By using \eqref{connection}, we can show that the maps
$[\nabla_i^\phi, \nabla_j^\phi]: TX^\phi \longrightarrow TX^\phi$
and ${[} \tilde\nabla_i^\phi, \tilde\nabla_j^\phi]: \tilde TX^\phi
\longrightarrow \tilde TX^\phi$ are left and right $(\cA,
\ast_\phi)$-module homomorphisms respectively. Namely, for all $Z\in
TX^\phi$, $\ W\in \tilde TX^\phi$ and $f\in\cA$,
\begin{eqnarray*}
[\nabla_i^\phi, \nabla_j^\phi](f\ast_\phi Z) &=& f\ast_\phi
[\nabla_i^\phi, \nabla_j^\phi]Z,\\
{[} \tilde\nabla_i^\phi, \tilde\nabla_j^\phi](W\ast_\phi
f)&=&[\tilde\nabla_i^\phi, \tilde\nabla_j^\phi]W\ast_\phi f.
\end{eqnarray*}
Thus we can define the curvatures ${}^\phi{R}_{i j k}^l$,
${}^\phi\tilde{R}_{i j k}^l$ as before by
\begin{eqnarray*}
[\nabla_i^\phi, \nabla_j^\phi]E_k^\phi &=&{\, }^\phi{R}_{k i j}^l\ast_\phi E_l^\phi, \\
{[}\tilde\nabla_i^\phi, \tilde\nabla_j^\phi]E_k^\phi &=&
E_l^\phi\ast_\phi{\, }^\phi\tilde{R}_{k i j}^l,
\end{eqnarray*}
and also construct their relatives such as ${}^\phi{R}_{i j k l}$,
${}^\phi\tilde{R}_{i j k l}$. Then ${}^\phi{R}_{i j k}^l$,
${}^\phi\tilde{R}_{i j k}^l$  are given by the formulae of Lemma
\ref{curvature1} with $\ast$ replaced by $\ast_\phi$, $\partial_i$
by $\partial^\phi_i$, $\Gamma$ by ${\, }^\phi\Gamma$ and
$\tilde\Gamma$ by ${\, }^\phi\tilde\Gamma$. Also, Proposition
\ref{curvature3} is still valid in the present case, and
${}^\phi{R}_{i j k}^l$, ${}^\phi\tilde{R}_{i j k}^l$ satisfy the
Bianchi identities (see Theorem \ref{BIs}).

\medskip

Now we examine properties of noncommutative surfaces under general
coordinate transformations. Let $X=(X^1, X^2, \dots, X^m)$ be an
element of $\cA^m$. We assume that $X$ gives rise to a
noncommutative surface over $(\cA, \ast)$.  Then we have the
following related noncommutative surfaces
\begin{eqnarray*}
\begin{aligned}
&\hat X&=&(\phi(X^1), \phi(X^2), \dots, \phi(X^m) ), &&
\text{over\ } \  (\cA, \ast),& \\
&X^\phi&=&(X^1, X^2, \dots, X^m), && \text{over\ } \  (\cA,
\ast_\phi),&
\end{aligned}
\end{eqnarray*}
associated with $X$. We call $X^\phi$ over $(\cA, \ast_\phi)$ the
{\em reparametrization} of the noncommutative surface $X$ over
$(\cA, \ast)$ in terms of $u=\phi^{-1}(t)$.
\begin{remark}
Note that $\cA^m$ is regarded as an $(\cA, \ast)$-module when we
study the noncommutative surface $X$, and regarded as an $(\cA,
\ast_\phi)$-module when we study $X^\phi$. Thus  even though $X$
and $X^\phi$ are the same element in $\cA^m$, they have quite
different meanings when the module structures of $\cA^m$ are taken
into account.
\end{remark}

Denote by $\hat g$ the metric, by $\hat{\Gamma}_{i j}^k$ and
$\hat{\tilde\Gamma}_{i j}^k$ the Christoffel symbols, and by
$\hat{R}_{i j k}^l$, $\hat{\tilde{R}}_{i j k}^l$ the Riemannian
curvatures of $\hat X$. They can be computed by using
$n$-dimensional generalizations of the relevant formulae derived in
Section \ref{surfaces}. The metric, curvature and other related
objects of the noncommutative surface $X^\phi$ over $(\cA,
\ast_\phi)$ are given in the last subsection.

\begin{theorem}\label{covariance}
There exist the following relations:
\begin{eqnarray*}
\begin{aligned}
&\gphi_{i j} = \phi^{-1}(\hat g_{i j}), & \quad& \gphi^{i j} =
\phi^{-1}\left(\hat{g}^{i j}\right), \\
&{\, }^\phi\Gamma_{i j}^k = \phi^{-1}(\hat\Gamma_{i j}^k), & \quad&
{\, }^\phi\tilde \Gamma_{i j}^k =\phi^{-1}(\hat{\tilde\Gamma}_{i
j}^k),\\
&{}^\phi{R}_{i j k}^l = \phi^{-1}(\hat{R}_{i j k}^l), &\quad&
{}^\phi\tilde{R}_{i j k}^l= \phi^{-1}(\hat{\tilde{R}}_{i j k}^l).
\end{aligned}
\end{eqnarray*}
\end{theorem}

\begin{remark}
We shall see in the next section that this theorem leads to the
standard transformation rules for the metric, connection and
curvature tensors in the commutative setting when we take the limit
$\barh\rightarrow 0$.
\end{remark}

\begin{proof}[Proof of Theorem \ref{covariance}]
Consider the first relation. Since $\phi^{-1}$ is an algebraic
isomorphism from $(\cA, \ast)$ to $(\cA, \ast_\phi)$, we have
\[
\phi^{-1}(\hat g_{i j})= \phi^{-1}(\partial_i\hat X\bullet\partial_j\hat X
)=\phi^{-1}(\partial_i\hat X)\bullet_\phi \phi^{-1}(\partial_j\hat
X ).
\]
Using $\phi^{-1}(\partial_i\hat X)=\partial^\phi_i X$, we obtain
\[
\phi^{-1}(\hat E_i)= E_i^\phi, \quad \forall i.
\]
Thus
\[
\phi^{-1}(\hat g_{i j})  =E_i^\phi\bullet_\phi E_j^\phi=\gphi_{i j}.
\]

Since $\phi$ maps $1$ to itself, it follows that $\phi^{-1}(\hat
g^{i j})=\gphi^{i j}$. Now
\begin{eqnarray*}
\phi^{-1}(\hat\Gamma^k_{i j}) &=& \phi^{-1}(\partial_i \hat
E_j\bullet
\hat E_l\ast\hat g^{l k} )\\
&=& \phi^{-1}(\partial_i \hat E_j)\bullet_\phi
\phi^{-1}(\hat E_l)\ast_\phi \phi^{-1}(\hat g^{l k} )\\
&=& \partial^\phi_i E^\phi_j\bullet_\phi E^\phi_l\ast_\phi
\gphi^{l k}\\ &=&{\, }^\phi\Gamma_{i j}^k.
\end{eqnarray*}
The other relations can also be proven similarly by using the fact
that $\phi^{-1}$ is an algebraic isomorphism. We omit the details.
\end{proof}

It is useful to observe how the covariant derivatives transform
under general coordinate transformations. We have
\begin{eqnarray*}
\nabla_i^\phi E_j^\phi &=& \partial_i^\phi  E_j^\phi + {\,
}^\phi\Gamma_{i j}^k\ast_\phi E_k^\phi\\
&=& \phi^{-1}(\partial_i  \hat E_j) + \phi^{-1}(\hat\Gamma_{i
j}^k\ast \hat E_k) \\
&=& \phi^{-1}(\hat\nabla_i\hat E_j),
\end{eqnarray*}
where $\hat\nabla_i$ is the covariant derivative in terms of the
Christoffel symbols $\hat\Gamma_{i j}^k$.

\begin{remark}\label{ast-phi}
The gauge transformation $\phi$ that procures the ``general
coordinate transformation" also changes the algebra $(\cA, \ast)$ to
$(\cA, \ast_\phi)$, thus inducing a map between noncommutative
surfaces defined over gauge equivalent noncommutative associative
algebras. This is very different from what happens in the
commutative case, but appears to be necessary in the noncommutative
setting.
\end{remark}

\begin{remark}\label{invariance}
Although the concept of covariance under the gauge transformation
$\phi$ is transparent and the considerations above show that our
construction of noncommutative surfaces is indeed covariant under
such transformations, it appears that the concept of invariance
becomes more subtle. In the classical case, a scalar is a function
on a manifold, which takes a value at each point of the manifold.
Invariance means that when we evaluate the function at ``the same
point" on the manifold, we get the same value (a real or complex
number) regardless of the coordinate system which we use for the
calculation. In the non-commutative case, elements of $\cA$ are not
numbers. When a general coordinate transformation is performed, the
algebraic structure of $\cA$ changes. It becomes rather unclear how
to compare elements in two different algebras.
\end{remark}

\subsection{Comparison with classical case}
One obvious question is the classical analogues of the differential
operators $\partial_i^\phi$ and the gauge transformations in
$\cG(\cA)$ which bring about the general coordinate transformations.
We address this question below. Morally, one should regard
$\cG(\cA)$ as the ``diffeomorphism group" of the surface, and
$\partial_i^\phi$ as ``differentiation" with respect to the new
coordinate $u^i:=\phi^{-1}(t^i)$.

Consider an element $\phi \in \cG(\cA)$. In the classical limit
(that is, $\barh=0$), $\phi$ obviously reduces to an element
$\exp(\nu)$ in the diffeomorphism group of $U$, where
$v=\epsilon^i(t)\partial_i$ is a smooth tangent vector field on $U$
classically. Then for any two smooth functions $a(t)$ and $b(t)$,
the Leibniz rule $v(a b) = v(a) b + a v(b)$ implies that $\exp(v)(a
b) = (\exp(v)a) (\exp(v) b)$ {at the classical level}. By regarding
a smooth function $a(t)$ as a power series in $t$ (or $t$ translated
by some constants) we then easily see that
\begin{eqnarray}\label{classical}
\exp(v)a(t) = a( \exp(v) t) && \text{at the classical level.}
\end{eqnarray}
Now for all $f(t)\in \cA$,
\begin{eqnarray*}
\partial_i^\phi \phi^{-1}(f(t)) &=&\phi^{-1}(\partial_i f(t)) =
\frac{\partial f(u)}{\partial u^i} \mod \barh\\
&=&\frac{\partial \phi^{-1}(f(t))}{\partial u^i} \mod \barh,
\end{eqnarray*}
where we have used \eqref{classical}. Replacing $f(t)$ by
$\phi(f(t))$ in the above computations we arrive at
\[
\partial_i^\phi f(t) = \frac{\partial f(t)}{\partial u^i} \mod
\barh.
\]
Using this result, we obtain
\[
\gphi_{i j} = \phi^{-1}\left( \phi\left(\frac{\partial
X(t)}{\partial u^i}\right) \cdot \phi\left(\frac{\partial
X(t)}{\partial u^j}\right)\right) \mod\barh,
\]
where the $\cdot$ on the right-hand side is the usual scalar product
for $\R^n$. Up to $\barh$ terms,
\[
\phi^{-1}\left( \phi\left(\frac{\partial X(t)}{\partial
u^i}\right) \cdot \phi\left(\frac{\partial X(t)}{\partial
u^j}\right)\right) = \frac{\partial X(t)}{\partial u^i} \cdot
\frac{\partial X(t)}{\partial u^j} \mod\barh,
\]
hence
\[
\gphi_{i j} = \frac{\partial t^p}{\partial u^i} g_{p
q}(t)\frac{\partial t^q}{\partial u^j}\mod\barh.
\]
This is the usual transformation rule for the metric if we ignore
terms of order $\ge 1$ in $\barh$.

It is fairly clear now that we shall also recover the usual
transformation rules for the Christoffel symbols and curvatures in
the classical limit $\barh\rightarrow 0$. We omit the proof.

\section{Noncommutative surfaces: sketch of general theory}\label{generalization}

In the earlier sections, we presented a theory of noncommutative
$n$-dimensional surfaces over a deformation of the algebra of smooth
functions on a region $U\subset\R^n$. This theory readily
generalizes to arbitrary associative algebras with derivations.
Below is a brief outline of the general theory.

Let $\cA$ be an arbitrary unital associative algebra over a
commutative ring $k$.  We shall write $a b$ as the product of any
two elements $a, b\in \cA$. Let $Z(\cA)$ be the center of $\cA$.
Then the set of derivations of $\cA$ forms a left $Z(\cA)$-module
such that for any derivation $d$ and $z\in Z(\cA)$, $z d$ is the
derivation which maps any $a\in\cA$ to $zd(a)$. We require $\cA$ to
have the following properties:
\begin{quote}
{\em the associative algebra $\cA$ has a set of mutually commutative
and $Z(\cA)$-linearly independent derivations $\partial_i$ ($i=1, 2,
\dots, n$)}.
\end{quote}
\begin{remark}
The Moyal algebra satisfies these conditions. However in general,
the assumptions impose stringent constraints on the noncommutative
algebras under consideration.
\end{remark}

Let $\cA^m$ be the free $\cA$-module of rank $m$.   Define a dot
product
\[
\bullet: \cA^m\otimes_k \cA^m \longrightarrow \cA, \quad A\bullet B=
a_i  b^i
\]
for any $A=(a_1, \dots, a_m)$ and $B=(b^1, \dots, b^m)$ in $\cA^m$.
Let $X=(X^1, X^2, \dots, X^m)$ be an element of $\cA^m$ for some
fixed $m>n$.  As before, we define
\begin{eqnarray}
E_i = \partial_i X = (\partial_i X^1, \partial_i X^2, \dots,
\partial_iX^m),
\end{eqnarray}
and construct an $n\times n$ matrix $g$ over $\cA$  with entries
\begin{eqnarray}
g_{i j} = E_i\bullet E_j.
\end{eqnarray}
We say that {\em $X$ defines a noncommutative surface over $\cA$ if
$g\in GL_n(\cA)$, and call $g$ the metric of the noncommutative
surface}.

Clearly the $Z(\cA)$-linear independence requirement on the
derivations is necessary in order for any invertible $g$ to exist.
If $g\in GL_n(\cA)$, then
\[
TX=\{z^i E_i \mid z^i\in\cA\}, \quad \tilde TX=\{E_i w^i \mid
z^i\in\cA\}
\]
are finitely generated projective (left or right)
$\cA$-modules, which are taken to be the tangent bundles of the
noncommutative surface. The metric defines a map
\[ g: TX\otimes_k \tilde TX \longrightarrow \cA,
\quad Z\otimes W \mapsto g(Z, W)=Z\bullet W \] of two-side
$\cA$-modules. We define connections
\[ \{ \nabla_i:
TX\longrightarrow TX \mid i=1, \dots, n\}, \quad \{\tilde\nabla_i:
\tilde TX\longrightarrow \tilde TX \mid i=1, \dots, n\}
\]
on the left and right tangent bundles respectively by generalizing
the standard procedure in the theory of surfaces \cite{doC}:
\begin{eqnarray}
\begin{aligned}
&\nabla_i (f Z) = (\partial_i f) Z + f \nabla_i Z, &\quad&
\forall f\in\cA, Z\in TX, \\
&g(\nabla_i E_j, E_l)= \partial_i E_j\bullet E_l;
\end{aligned}
\end{eqnarray}
and
\begin{eqnarray}
\begin{aligned}
&\tilde\nabla_i (W f) =W \partial_i f  + \tilde\nabla_i W  f,
&\quad&
\forall f\in\cA, W\in \tilde TX, \\
&g(E_j, \tilde\nabla_i E_l)= E_j\bullet\partial_i  E_l.
\end{aligned}
\end{eqnarray}
Then the connections are compatible with the metric in the sense
of Proposition \ref{compatible}.

It can be shown that $[\nabla_i, \nabla_j]: TX\longrightarrow TX$
and $[\tilde\nabla_i, \tilde\nabla_j]: \tilde TX\longrightarrow
\tilde TX$ are left and right $\cA$-module homomorphisms
respectively. Thus one can define curvatures of the connections on
the left and right tangent bundles in the same way as in Sections
\ref{surfaces} and \ref{general} (see equation \eqref{curvature}).
We shall not present the details here, but merely point out that the
various curvatures still satisfy Propositions \ref{curvature3} and
\ref{BIs}.

\section{Discussion and conclusions}

Riemannian geometry is the underlying structure of Einstein's theory
of general relativity, and historically the realization of this fact
led to important further developments. In this paper we have
developed a Riemannian geometry of noncommutative surfaces as a
first step towards the construction of a consistent noncommutative
gravitational theory.

Our treatment starts from the simplest nontrivial examples, on which
the general theory is gradually elaborated. We begin by constructing
a noncommutative Riemannian geometry for noncommutative analogues of
2-dimensional surfaces embedded in 3-space, working over an
associative algebra $\cA$, which is a deformation of the algebra of
smooth functions on a region of $\R^2$. On $\cA^3$ we define a
``dot-product" analogous to the usual scalar product for the
Euclidean 3-space.  An embedding $X$ of a noncommutative surface is
defined to be an element of $\cA^3$ satisfying certain conditions.
Partial derivatives of $X$ then generate a left and also a right
projective $\cA$-module, which are taken to be the tangent bundles
of the noncommutative surface. Now the dot-product on $\cA^3$
induces a metric on the tangent bundles, and connections on the
tangent bundles can also be introduced following the standard
procedure in the theory of surfaces \cite{doC}.  Much of the
classical differential geometry for surfaces is shown to generalize
naturally to this noncommutative setting. We point out that the
embeddings greatly help the understanding of the geometry of
noncommutative surfaces.

From the noncommutative Riemannian geometry of the 2-dimensional
surfaces we go straightforwardly to the generalization to
noncommutative geometries corresponding to $n$-dimensional surfaces
embedded in spaces of higher dimensions. In higher dimensions, the
Riemannian curvature becomes much more complicated, thus it is
useful to know its symmetries. A result on this is the
noncommutative analogues of Bianchi identities proved in Theorem
\ref{BIs}.

We also observe that there exists another object $\Theta_j^i$ (see
\eqref{Theta}), which is distinct from the Ricci curvature $R^i_j$
but also reduces to the classical Ricci curvature in the commutative
case. Contracting indices in the second noncommutative Bianchi
identity, we arrive at an equation involving ``covariant
derivatives" of both $R^i_j$ and $\Theta_j^i$. This appears to
suggest that Einstein's equation acquires modification in the
noncommutative setting, as shown in Subsection \ref{sect:Einstein}.
Work along this line is in progress \cite{CTZZ}.

A special emphasis is put on the covariance under general coordinate
transformations, as the fundamental principle of general relativity.
It is physically natural that under general coordinate
transformations, the frame-dependent Moyal star-product would
change. In this spirit, we introduce in Section
\ref{transformations} general coordinate transformations for
noncommutative surfaces, in the form of gauge transformations on the
underlying noncommutative associative algebra $\cA$, which change as
well the multiplication of the underlying associative algebra $\cA$,
turning it into another algebra nontrivially isomorphic to $\cA$. By
comparison with classical Riemannian geometry, we show that the
gauge transformations should be considered as noncommutative
analogues of diffeomorphisms. We emphasize that in our construction
we allow for all possible diffeomorphisms, and not only those
preserving the $\theta$-matrix constant, as has been done so far in
most of the literature in the field.

The results are eventually generalized to a theory of noncommutative
Riemannian geometry of $n$-dimensional surfaces over unital
associative algebras with derivations. This is outlined in Section
\ref{generalization}. Noncommutative surfaces should provide a
useful test ground for generalizing Riemannian geometry to the
noncommutative setting.

From the point of view of physics, noncommutative surfaces with
Minkowski signature, which were briefly alluded to in Remark
\ref{pseudo}, are more interesting. To treat such noncommutative
surfaces in depth, more care will be required. It is known in the
commutative case that the realization of a pseudo-Riemannian surface
in the flat Minkowski space may contain isotropic subsets with
singular metrics.

The ultimate aim is to obtain a noncommutative version of
gravitational theory, covariant under appropriately defined general
coordinate transformations and, possibly, compatible with the
gauging of the twisted Poincar\'e symmetry \cite{CKNT,CKNT1}, in analogy with the
classical works of Utiyama \cite{uti} and Kibble \cite{kibble} (for a recent attempt in this direction, see \cite{COTZ}).

\bigskip

\noindent{\bf Acknowledgement}: We are much indebted to A. H. Chamseddine, J. Gracia-Bond\'ia,
J. Madore and G. Zet for useful discussions and valuable suggestions
on the manuscript. Partial financial support from the Australian
Research Council, National Science Foundation of China (grants
10725105, 10731080), NKBRPC (2006CB805905), the Chinese Academy of
Sciences, and the Academy of Finland (grant 121720)  is gratefully
acknowledged.


\begin{thebibliography}{9999}

\bibitem{Co} A. Connes, {\em Noncommutative geometry}. Academic Press (1994).

\bibitem{GVF} J. M. Gracia-Bond\'ia, J. C. V\'arilly, H. Figueroa,  Elements of
noncommutative geometry. Birkh\"auser Advanced Texts: Basler
Lehrb\"ucher.  Birkh\"auser Boston, Inc., Boston, MA (2001).

\bibitem{SV} J. T. Stafford, M. Van den Bergh, Noncommutative curves and
noncommutative surfaces. Bull. Amer. Math. Soc. (N.S.) {\bf 38}
171 (2001).

\bibitem{Ko} M. Kontsevich, Deformation quantization of Poisson manifolds.
Lett. Math. Phys.  {\bf 66}  157 (2003), q-alg/9709040.

\bibitem{SW} N. Seiberg,  E. Witten,  String theory and
noncommutative geometry.  JHEP  {\bf 9909} 032 (1999), hep-th/9908142.

\bibitem{DN} M.R. Douglas,  N. A. Nekrasov, Noncommutative field theory.
Rev. Mod. Phys.  {\bf 73} 977 (2001), hep-th/0106048.

\bibitem{V} J. C.  V\'arilly,  An introduction to noncommutative geometry. EMS
Series of Lectures in Mathematics. European Mathematical Society
(EMS), Z\"urich, 2006.

\bibitem{DFR} S. Doplicher, K. Fredenhagen, J. E. Roberts,
The quantum structure of spacetime at the Planck scale and quantum
fields. Commun. Math. Phys.  {\bf 172} 187 (1995), hep-th/0303037.

\bibitem{ADMW1}  P. Aschieri, C. Blohmann, M. Dimitrijevic, F. Meyer, P. Schupp, J. Wess,
A gravity theory on noncommutative spaces.  Class. Quant. Grav. {\bf
22} 3511 (2005), hep-th/0504183.

\bibitem{ADMW2} P. Aschieri, M. Dimitrijevic, F. Meyer, J. Wess, Noncommutative
geometry and gravity. Class. Quant. Grav. {\bf 23} 1883
(2006), hep-th/0510059.

\bibitem{BGMZ} M. Buric, T. Grammatikopoulos, J. Madore, G. Zoupanos,
Gravity and the structure of noncommutative algebras. JHEP {\bf 0604} 054
(2006), hep-th/0603044.

\bibitem{C01} A. H. Chamseddine, Complexified gravity in noncommutative spaces.
Commun. Math. Phys. {\bf 218}  283 (2001), hep-th/0005222.

\bibitem{C04} A. H. Chamseddine, $SL(2,\C)$ gravity with a complex vierbein and its noncommutative
extension. Phys. Rev. {\bf D 69} 024015 (2004), hep-th/0309166.

\bibitem{MM} J. Madore, J. Mourad, Quantum space-time and classical
gravity. J. Math. Phys. {\bf 39} 423 (1998), gr-qc/9607060.

\bibitem{M05}  S. Majid, Noncommutative Riemannian and spin geometry of the standard $q$-sphere.
Commun. Math. Phys.  {\bf 256} 255  (2005).

\bibitem{CKNT} M. Chaichian, P.P. Kulish, K. Nishijima, A. Tureanu,  On a
Lorentz-invariant interpretation of noncommutative space-time and
its implications on noncommutative QFT. Phys. Lett. {\bf B 604} 98
(2004), hep-th/0408069.

\bibitem{CKNT1}
M. Chaichian, P. Pre\v{s}najder, A. Tureanu, New concept of
relativistic invariance in noncommutative space-time: Twisted
Poincare symmetry and its implications. Phys. Rev. Lett. {\bf 94}
151602 (2005), hep-th/0409096.

\bibitem{AMV} L. \'Alvarez-Gaum\'e, F. Meyer,  M. A. Vazquez-Mozo,
Comments on noncommutative gravity. Nucl. Phys. {\bf B 75} 392
(2006), hep-th/0605113.

\bibitem{CT} M. Chaichian, A. Tureanu,  Twist symmetry and gauge invariance.
Phys. Lett. {\bf B 637} 199 (2006), hep-th/0604025.

\bibitem{CTZ}
M. Chaichian, A. Tureanu, G. Zet,  Twist as a symmetry principle and
the noncommutative gauge theory formulation. Phys. Lett. {\bf B651} 319
(2007), hep-th/0607179.

\bibitem{doC} M. P. do Carmo, {\em Differential geometry of curves and surfaces}.
              Englewood Cliffs, N.J. : Prentice-Hall, 1976.

\bibitem{DFN} B. A. Dubrovin, A. T. Fomenko and S. P. Novikov,
Modern geometry - methods and applications: Part II. The geometry
and topology of manifolds. Springer-Verlag, Heidelberg, 1985.

\bibitem{Ge} M. Gerstenhaber, On the deformation of rings and algebras, Ann. Math. (2)
{\bf 79} 59 (1964).


\bibitem{DHLS} L. Dabrowski, P. M. Hajac, G. Landi, P. Siniscalco, Metrics and
pairs of left and right connections on bimodules.  J. Math. Phys.
{\bf 37} 4635 (1996).

\bibitem{DM} M. Dubois-Violette,  P. W. Michor, Connections on central
bimodules in noncommutative differential geometry.  J. Geom. Phys.
{\bf 20} 218 (1996).

\bibitem{DMMM} M. Dubois-Violette, J. Madore, T. Masson, J.
Mourad. On curvature in noncommutative geometry. J. Math. Phys. {\bf
37} 4089 (1996).

\bibitem{GLMV} J. M. Gracia-Bond\'ia, F. Lizzi, G. Marmo, P. Vitale, Infinitely many
star products to play with. JHEP {\bf 0204} 026 (2002), hep-th/0112092.

\bibitem{Z} A. Zotov,  On relation between Moyal and
Kontsevich quantum products: direct evaluation up to the
$\hbar^3$-order. Mod. Phys. Lett. {\bf A16} 615 (2001), hep-th/0007072.

\bibitem{GGR} V. Gayral, J.M. Gracia-Bond\'ia,
F. Ruiz Ruiz, Position-dependent noncommutative products: Classical
construction and field theory. Nucl. Phys. {\bf B727}
513 (2005), hep-th/0504022.

\bibitem{CTZZ}
M. Chaichian, A. Tureanu, R. B. Zhang, X. Zhang, Work in progress.

\bibitem{uti}
R. Utiyama, Invariant theoretical interpretation of interaction,
Phys. Rev. {\bf 101} 1597 (1956).

\bibitem{kibble}
T. W. B. Kibble, Lorentz invariance and the gravitational field, J.
Math. Phys. {\bf 2} 212 (1961).

\bibitem{COTZ}
M. Chaichian, M. Oksanen, A. Tureanu, G. Zet, Gauging the twisted Poincar\'e symmetry as noncommutative theory of gravitation, arXiv:0807.0733 [hep-th].


\end{thebibliography}
\end{document}